\documentclass[12pt,amssymb]{article} 

\usepackage{mathrsfs}
\usepackage{amssymb}
\usepackage[dvips]{graphicx}

\newcommand{\newsection}[1]{
\addtocounter{section}{1} \setcounter{equation}{0}
\setcounter{subsection}{0} \addcontentsline{toc}{section}{\protect
\numberline{\arabic{section}}{{\rm #1}}} \vglue .6cm \pagebreak[3]
\noindent{ \bf  \thesection. #1}\nopagebreak[4]\par\vskip .3cm}
\newcommand{\newsubsection}[1]{
\addtocounter{subsection}{1}\setcounter{subsubsection}{0}
\addcontentsline{toc}{subsection}{\protect
\numberline{\arabic{section}.\arabic{subsection}}{#1}} \vglue .4cm
\pagebreak[3] \noindent{\it \thesubsection.
#1}\nopagebreak[4]\par\vskip .3cm}

\makeatletter
\newcommand{\seclabel}[1]{%
  \@bsphack
  \protected@write\@auxout{}%
     {\string\newlabel{#1}{{\thesection}{\thepage}}}
  \@esphack
  }
\newcommand{\subseclabel}[1]{%
  \@bsphack
  \protected@write\@auxout{}%
     {\string\newlabel{#1}{{\thesubsection}{\thepage}}}
  \@esphack
  }
\newcommand{\tablabel}[1]{%
  \@bsphack
  \protected@write\@auxout{}%
     {\string\newlabel{#1}{{\arabic{tabnum}}{\thepage}}}
  \@esphack
  }
\makeatother

\renewcommand{\theequation}{\thesection.\arabic{equation}}

\newlength{\extraspace}
\newlength{\extraspaces}
\newcounter{dummy}
\newcommand{\bc}{\begin{center}}
\newcommand{\ec}{\end{center}}
\newcommand{\be}{\begin{equation}
\addtolength{\abovedisplayskip}{\extraspaces}
\addtolength{\belowdisplayskip}{\extraspaces}
\addtolength{\abovedisplayshortskip}{\extraspace}
\addtolength{\belowdisplayshortskip}{\extraspace}}
\newcommand{\ee}{\end{equation}}

\newcommand{\ba}{\begin{eqnarray}
\addtolength{\abovedisplayskip}{\extraspaces}
\addtolength{\belowdisplayskip}{\extraspaces}
\addtolength{\abovedisplayshortskip}{\extraspace}
\addtolength{\belowdisplayshortskip}{\extraspace}}
\newcommand{\ea}{\end{eqnarray}}

\newcommand{\is}{& \!\! = \!\! &}
\newcommand{\ban}{\begin{eqnarray*}
\addtolength{\abovedisplayskip}{\extraspaces}
\addtolength{\belowdisplayskip}{\extraspaces}
\addtolength{\abovedisplayshortskip}{\extraspace}
\addtolength{\belowdisplayshortskip}{\extraspace}}
\newcommand{\ean}{\end{eqnarray*}}
\newcommand{\baa}{
\addtocounter{equation}{1} \setcounter{dummy}{\value{equation}}
\setcounter{equation}{0}
\renewcommand{\theequation}{\thesection.\arabic{dummy}\alph{equation}}
\begin{eqnarray}
\addtolength{\abovedisplayskip}{\extraspaces}
\addtolength{\belowdisplayskip}{\extraspaces}
\addtolength{\abovedisplayshortskip}{\extraspace}
\addtolength{\belowdisplayshortskip}{\extraspace}}
\newcommand{\eaa}{
\end{eqnarray}
\setcounter{equation}{\value{dummy}}
\renewcommand{\theequation}{\thesection.\arabic{equation}}}

\input epsf

\newcounter{fignum}
\newcounter{tabel}

\newcounter{tabnum}
\newcommand{\vev}[1]{\left\langle #1\right\rangle}
\newcommand{\ket}[1]{\left| #1 \right\rangle}
\newcommand{\bra}[1]{\left\langle #1 \right|}
\newcommand{\half}{\frac{1}{2}}
\newcommand{\del}{\partial}

\newcommand{\eol}{\nonumber \\}

\newcommand{\Hom}{{\rm Hom}}

\newcommand{\bt}{{\bf 10}}
\newcommand{\bfv}{{\bf 5}}
\newcommand{\bfb}{{\overline{\bf 5 \!}\,}}
\newcommand{\btb}{{\overline{\bf 10 \!}\,}}

\begin{document}

%
%

\begin{flushright}
May 2009\\
AEI-2009-047
\end{flushright}
\vspace{2cm}

\thispagestyle{empty}

%
%

\begin{center}
{\Large\bf  Hitchin's Equations and $M$-Theory Phenomenology
 \\[13mm] }

{\sc Tony Pantev}\\[2.5mm]
{\it Department of Mathematics, University of Pennsylvania \\
Philadelphia, PA 19104-6395, USA}\\[9mm]

{\sc Martijn Wijnholt}\\[2.5mm]
{\it Max Planck Institute (Albert Einstein Institute)\\
Am M\"uhlenberg 1 \\
D-14476 Potsdam-Golm, Germany }\\
[30mm]

 {\sc Abstract}

\end{center}

Phenomenological compactifications of $M$-theory involve
$7$-manifolds with $G_2$ holonomy and various singularities. Here we
study local geometries with such singularities, by thinking of them
as compactifications of $7d$ supersymmetric Yang-Mills theory on a
three-manifold $Q_3$. We give a general discussion of
compactifications of $7d$ Yang-Mills theory in terms of Higgs
bundles on $Q_3$. We show they can be constructed using spectral
covers, which are Lagrangian branes with a flat connection in the
cotangent bundle $T^*Q_3$. We explain the dictionary with ALE
fibrations over $Q_3$ and conjecture that these configurations have
$G_2$ holonomy. We further develop tools to study the low energy
effective theory of such a model. We show that the naive massless
spectrum is corrected by instanton effects. Taking the instanton
effects into account, we find that the massless spectrum and many of
the interactions can be computed with Morse theoretic methods.

\newpage

\renewcommand{\Large}{\normalsize}

\tableofcontents

\newpage

\newsection{Introduction}

String theory vacua are explicit realizations of the ideas of Kaluza
and Klein on extra dimensions. As such, geometric structures
inevitably play a central role in studying such vacua. Geometric
techniques are widely used for phenomenological model building in
the heterotic string.

In the past few years we have learnt to start applying techniques
from geometric engineering to model building problems in type II
settings. Using ideas about exceptional collections, this led to the
construction of `local models' in which the force of gravity may be
treated as a small perturbation
\cite{Aldazabal:2000sa,Berenstein:2001nk,Verlinde:2005jr}. More
recently it led to the first new class of models since the
appearance of the heterotic string that can successfully explain
unification, namely $F$-theory GUTs
\cite{Donagi:2008ca,Beasley:2008dc,Hayashi:2008ba}. The $F$-theory
models allow for scenarios not available in the heterotic string,
such as gauge mediation, and hence may display some strikingly
different signatures. Since they have been relatively little
studied, it is desirable to develop the phenomenology and
mathematics of these models in more detail. A number of interesting
papers have recently appeared on this topic.

In this project we follow a slightly different direction. In light
of the line of research mentioned above, it is natural to ask if a
similar set of ideas based on geometric engineering may also be used
to address some of the difficulties encountered in constructing
phenomenological models in other type II settings. Here we would
like to address the construction of GUT models in $M$-theory.
Although such models were expected to exist and qualitative features
have been studied assuming their existence, there are currently no
examples or techniques for explicitly constructing them. For a
recent summary see \cite{Acharya:2008zi}.

One of the main ideas that we employ, in physical terms, is the
central role played by the BPS equations of a `worldvolume'
supersymmetric Yang-Mills theory, which lives in seven dimensions in
the present case. In mathematical terms we are dealing with Higgs
bundles, that is a gauge field and an adjoint Higgs field satisfying
a version of Hitchin's equations, which are exactly the BPS
equations of the $7d$ Yang-Mills theory. Even though we are in a
non-perturbative regime of string theory, the gauge theory
description can be trusted as long as the gauge and Higgs field are
slowly varying. This strategy was successfully used in recent
phenomenological constructions in $F$-theory, and we will see how it
carries over in $M$-theory.

On the other hand, supersymmetric compactifications of $M$-theory to
four dimensions are known to be given by $G_2$-manifolds. The
principle that the worldvolume gauge theory completely determines
the local geometry of the brane is well-established in other
contexts in string theory. Therefore we expect that our approach
should establish the existence of a large class of non-compact $G_2$
manifolds with singularities. We will outline how one recovers the
data of a non-compact $G_2$-manifold with singularities from the
data of the compactified $7d$ gauge theory, but we will not give a
complete proof of the correspondence in this paper. At any rate,
string, $M$ or $F$-theory only plays a secondary role in our
philosophy. The primary object of interest is the higher dimensional
Yang-Mills theory, and the main role of string/$M$/$F$-theory is to
provide a UV completion of this Yang-Mills theory.

Spectral covers are a powerful technique for constructing solutions
to Hitchin's equations. In the present setting, spectral covers
correspond to Lagrangian $A$-branes in the auxiliary Calabi-Yau
geometry $T^*S^3$, and thus much of our intuition about intersecting
Lagrangians can be carried over. To some extent this is not
surprising, because if we do not use exceptional gauge groups, then
we should be able to take a perturbative IIa limit and end up with a
configuration of $D6$-branes and orientifold planes on a Calabi-Yau.
However our Lagrangian branes are intrinsic to the $7d$ Yang-Mills
theory and do not assume any string or $M$-theory.

For phenomenological purposes it is important to understand the
spectrum and interactions in such models. Some qualitative results
have already been obtained in the literature. Here we will find that
these results may be better understood and extended using Morse
theory as a principal tool. As a result, we find that the massless
spectrum and many of the interactions reflect topological properties
of the configuration, and can be computed without any knowledge of
the solutions of the $D$-term equations. This is a remarkable
simplification which should be of great help in understanding the
phenomenological signatures of these models.

In this paper we mostly focus on abelian examples, although we will
make statements that apply more generally. Work on non-abelian
examples is still in progress \cite{PW}.

\newpage

\newsection{Local models in $M$-theory}

As mentioned in the introduction, our goal will be to construct
compactifications of the $7d$ supersymmetric Yang-Mills theory. Such
compactifications are mathematically described by Higgs bundles. In
order to explain the relevance of such compactifications to
$M$-theory, we have to explain how this data is related to $G_2$
manifolds with singularities. The main purpose of this section is to
set up the dictionary between ALE-fibrations in $M$-theory, Higgs
bundles, and spectral covers in an auxiliary Calabi-Yau geometry.

\newsubsection{General properties of $G_2$-manifolds}
\subseclabel{G2manifolds}

 We are interested in engineering
effectively four-dimensional models from $M$-theory. Since
$M$-theory lives in $11d$, this means we must compactify on a
seven-dimensional internal space $X_7$. Furthermore, if we require
$N=1$ supersymmetry in four dimensions, then $X_7$ should admit a
Killing spinor. As is well-known, the existence of a Killing spinor
implies that $X_7$ must admit a metric of $G_2$-holonomy. Such
metrics are hard to find explicitly. However, much like Calabi-Yau
metrics which are also hard to find, one may reformulate the problem
of finding $G_2$-metrics in terms of anti-symmetric tensor fields,
which are much easier to work with. For $G_2$ manifolds, the
relevant tensor field is a three-form $\Phi$.

Given a smooth 7-manifold, a three-form $\Phi$ is said to be stable
if it lies in an open orbit of $Gl(7)$. In terms of a 7-bein, $\Phi$
may be written as
\be \Phi = e^1 \wedge e^2 \wedge e^3 + e^i \wedge \Sigma_i \ee
where
\be \Sigma_1 = e^4 \wedge e^5 - e^6\wedge e^7, \qquad
 \Sigma_2 = e^4 \wedge e^6 - e^7\wedge e^5, \qquad
 \Sigma_3 = e^4 \wedge e^7 - e^5\wedge e^6
 \ee
From such a three-form, one may reconstruct a metric through the
following formulae \cite{HitchinG2}:
\ba g_{ij} &=& \det(s_{ij})^{-1/9} s_{ij} \eol s_{ij} \is -{1\over
144}\Phi_{i\mu_1\mu_2}\Phi_{j\mu_3\mu_4} \Phi_{\mu_5\mu_6\mu_7}
\epsilon^{\mu_1 \ldots \mu_7} \sim  -{1\over 144} \Phi_i\wedge
\Phi_j \wedge \Phi \ea
In terms of $\Phi$, the condition that the metric has $G_2$ holonomy
is equivalent to \cite{FernandezGray}
\ba\label{G2condition} d\Phi=0 &&\qquad ({\rm F-term})\eol  d*_\Phi\Phi=0 &&\qquad
({\rm D-term})\ea
The $*$-operator depends on the metric and hence implicitly on
$\Phi$, as we have indicated.

 These two equations can be obtained as
the critical points of two functionals which have a natural
four-dimensional interpretation. The first equation in
(\ref{G2condition}) is the equation of motion of a Chern-Simons
functional $W \sim \int \Phi \wedge d\Phi$. In fact, $M$-theory not
only yields a metric but also a three-form tensor field $C_{3}$, and
in $N=1$ SUSY compactifications it is natural to combine them in a
single complex three-form field $C_3 + i \Phi$. A quick way to see
this is by reducing on a circle to type IIa, in which case we get
the complexified K\"ahler form $B + iJ$ as required by
supersymmetry.
Including the dynamics of $C_3$, the BPS equations are generalized to
\ba\label{G2BPS} d(C_3 + i\Phi)=0  \eol d*_\Phi\Phi=0 \ea
and the Chern-Simons functional may be generalized to: %
\be\label{G2CS} W = {1\over 16\pi^2} \int_{X_7} (C_3 + i \Phi)
\wedge d (C_3 + i \Phi) \ee
This expression combines both the action of
\cite{Beasley:2002db,Gukov:1999gr,Acharya:2000ps} and that of
Hitchin \cite{HitchinStable}. It is interpreted as a term in the
four-dimensional superpotential.

The second equation in (\ref{G2BPS}) can be interpreted as a moment
map condition. The three-form field in $M$-theory transforms under a
group $\mathscr{G}$ of gauge transformations as $C_3 \to C_3 +
d\Lambda$. We also have a natural K\"ahler form associated to the
metric
\be \int_{X_7} (C_3' + i \Phi') \wedge \overline{*_\Phi ({C_3'' + i
\Phi''})} \ee
on the space of solutions of the $F$-terms. Then $i\,d^\dagger\Phi$
is the moment map associated to $\mathscr{G}$, and the second
equation in (\ref{G2BPS}) describes the critical points of a
$D$-term potential. Under a suitable stability condition, we would
expect that for every solution to the $F$-terms there exists a
unique solution to the $D$-terms in the same $\mathscr{G}^c$ orbit.
This is often guaranteed by the Kempf-Ness theorem, but the standard
version of this theorem does not apply here.

The linearized deformations of (\ref{G2BPS})  modulo gauge
transformations by $\mathscr{G}$ are counted by harmonic
three-forms, and these deformations are unobstructed
\cite{JoyceG2I}. Therefore the number of complex deformations of the
$G_2$ structure is given by $h^3(X_7)$. The K\"ahler potential on
moduli space turns out to be given by \cite{Beasley:2002db}

\be {\cal K} =  -3 \log\, {1\over 2\pi^2}{1\over 7} \int |C_3 + i
\Phi|^2 
\ee
In this paper we will be taking a slightly different point of view
however, and we will not be using these expressions explicitly.

To summarize, the main point of this subsection is that the
three-form $\Phi$ is an equivalent but more useful variable than the
$G_2$ metric itself. Moreover, in terms of these variables the
equations naturally split up into a set of first order equations
which can be interpreted as $F$-terms and $D$-terms.

\newsubsection{ALE fibrations}

\subseclabel{ALEfibrations}

Now consider a local $G_2$ manifold $X_7$ which is an ALE fibration
over a three-manifold $Q_3$. We will usually assume that the ALE
fibration has a section, and also use $Q_3$ to denote this section.
On each ALE fiber, there is a natural set of two-cycles $\alpha_i
\in H_2(ALE,{\bf Z})$ which intersect according to the Cartan matrix
associated to the ALE, generating an ADE root lattice $\Lambda$.
There is also a dual set of two-forms $\omega^i$.

The moduli space ${\cal M}(\Lambda)$ of $M$-theory on an ALE surface
is described as follows. Given a hyperk\"ahler structure $\{ {\cal
I,J,K} \}$ on the ALE, we can construct a triplet of two-forms
$\vec{\Omega} = ({\cal I}_\mu^\nu g_{\nu\lambda},{\cal J}_\mu^\nu
g_{\nu\lambda},{\cal K}_\mu^\nu g_{\nu\lambda})$. Their periods over
the $\alpha_i$ are the parameters
\be \int_{\alpha_i} \vec{\Omega} = \vec{\phi}_i, \ee
which describe complex structure and K\"ahler moduli of the ALE.
They are often called the FI parameters because they appear as such
in the hyperk\"ahler quotient construction of the ALE. They
naturally transform as a vector under an $SO(3)_R$ symmetry. In
addition, we may expand the $M$-theory three-form in terms of the
$\omega^i$, yielding $n$ vectors in seven dimensions, where $n$ is
the rank of the lattice $\Lambda$. Since the ALE preserves half of
the 32 supersymmetry generators, we are guaranteed to recover their
fermionic superpartners as well. In fact they are given by the same
internal wave-functions on the ALE. So for large sizes of the
vanishing cycles we get a supersymmetric $7d$ gauge theory with
gauge group $U(1)^n$. But what happens when the vanishing cycles are
small?

There are additional supersymmetric states obtained from wrapping
$M2$-branes on the vanishing cycles of the ALE. Their masses are
given by $m \sim m_{pl} |N^i\vec{\phi}_i|$, i.e. they are
proportional to the size of the vanishing cycle $N^i\alpha_i$ that
the membrane is wrapping. Quantizing such a particle yields a vector
multiplet, since this is the only non-gravitational multiplet
available in a $7d$ theory with $16$ supercharges. Since membranes
couple to $C_3$ in $11d$, the $U(1)^n$ charges of these states are
precisely those of the $W$-bosons of a non-abelian $ADE$ gauge
theory with root lattice $\Lambda$. Therefore the effective low
energy dynamics of $M$-theory on an ALE surface with small periods
should be described by the corresponding $7d$ supersymmetric
non-abelian ADE gauge theory. Additional evidence for this statement
can be obtained through heterotic/type II duality.

When we further fiber the ALE over $Q_3$, additional supersymmetries
will be broken. In $G_2$ compactifications, supersymmetry requires
that $C_3$ and $\Phi$ are paired into a complex three-form.
Expanding in a basis of exceptional cycles of the ALE, locally we
get $n$ complex one-forms on $Q_3$:
\be \int_{\alpha_j} C_3 + i \Phi = A_j + i\,\phi_j \ee
Since $\Phi$ describes zero modes of the metric, the one-form
$\phi_j$ must be identified with the triplet of adjoint scalars of
the $7d$ gauge theory encountered above. One can see this more
explicitly from the canonical expression of $\Phi$ in terms of a
7-bein. Therefore the three adjoint scalars associated to each
$\alpha_j$ must be twisted to a one-form on $Q_3$
\cite{Acharya:1998pm}. In other words, in order to preserve $N=1$
supersymmetry in four-dimensions, the $SO(3)_R$-symmetry acting on
the $\vec{\phi}$ is identified with the (dual of the) $SO(3)_Q$
structure group of $Q_3$. Equivalently, there exists a covariantly
constant tensor $J_\mu^\nu$ on $Q_3$ where $\mu$ transforms as a
vector under $SO(3)_Q$ and $\nu$ transforms as a vector of
$SO(3)_R$. In suitable coordinates we may write $J_\mu^\nu =
i\delta_\mu^\nu$.

Thus locally the ALE fibration may be described by a set of $n$
complex one-forms. There is an additional symmetry however which may
be used in gluing local patches together in a global model. Namely
the ALE has a diffeomorphism symmetry group which acts on the cycles
as the ADE Weyl group. Eg. for $A_{n-1}$ ALEs this is the symmetric
group $S_n$ on $n$ letters. This symmetry may be thought of as a
residual gauge symmetry from the non-abelian gauge theory. Thus
altogether we see that an ALE fibration may be described by $n$
one-forms on $Q_3$, with branch points across which the one-forms
may be permuted. Said differently, locally on $Q_3$ we have a map
from $Q_3$ into the parameter space
$\mathfrak{h}_{\mathbb{R}}^{\otimes 3}/W$ of the universal unfolding
of our ADE singularity. This is the essence of a Higgs bundle, as we
will explain in more detail in section \ref{SpectralCover}.

Conversely, given a configuration for the $\phi_i$ on $Q_3$, we may
try to reconstruct an ALE fibration over $Q_3$ with $G_2$ holonomy.
To first order we get
\be\label{G2series} \Phi = \Phi_0 + {\phi}_i \wedge {\omega}^i +
\ldots\ee
where $\Phi_0$ corresponds to the three-form for a constant $ALE$
fibration, which certainly exists (it may be written down
explicitly). The equations $d\Phi = d*\Phi = 0$ put constraints on
the $\phi_i$ and on the higher order terms. By analogy with
Kodaira-Spencer theory for Calabi-Yau manifolds
\cite{TianCY,TodorovCY}, we conjecture that if the $\phi_i$ satisfy
certain first order equations discussed below as well as tadpole
constraints, and if $Q_3$ has non-negative curvature (so as to avoid
curvature singularities at finite distance from the zero section
that one would otherwise likely have), then the above series may be
uniquely completed and has a finite radius of convergence.

So our main point here is that the data of the ALE fibration may be
described as a field configuration in a supersymmetric $7d$
Yang-Mills theory living on $Q_3$. This naturally leads to a set of
equations which we would expect the data to satisfy, in the limit
that the fields are slowly varying.

\newsubsection{Hitchin system}

The BPS conditions in the $7d$ gauge theory are given by reducing
the BPS conditions for ten-dimensional YM theory to seven
dimensions. In $10d$ we have
\be \delta \lambda = {\sf F}_{\mu\nu}\,\Gamma^{\mu\nu}\epsilon =0
\ee
To preserve $3+1d$ Poincar\'e invariance, we assume only field
configurations on the $6d$ internal space are turned on. On a
Calabi-Yau manifold, this yields the well-known Hermitian-Yang-Mills
equations
\be {\sf F}^{2,0} = 0, \qquad g^{\mu\bar{\nu}} {\sf
F}_{\mu\bar{\nu}}=0 \ee
Next we reduce these equations to $3d$. This is well-understood
\cite{HitchinHiggs,Leung:2000zv}.

Let us work on a local patch of $Q_3$. Then we can put a Calabi-Yau
metric on the tangent bundle which is semi-flat. That is, the metric
on the tangent bundle is expressed as $ds^2 = K_{ij}(x)(dx^i dx^j +
dy^i dy^j)$, where $K$ is a real K\"ahler potential which satisfies
a real Monge-Amp\`ere equation, and $x$ and $y$ are coordinates
along $Q_3$ and along the bundle directions respectively. There is a
natural complex structure in which the complex coordinates are given
by $z^j = x^j + i y^j$. To perform the reduction, we assume that the
gauge field is independent of the $y$ coordinates, and we write the
gauge field on the tangent bundle as
\be {\sf A} =  A_i(x) dx^i + \phi_j(x) dy^j = (A + \phi_J)_i dx^i
\ee
where we used $dy^j = J^j_i dx^i$ and wrote $\phi_{Ji} = \phi_j
J^j_i\sim \sqrt{-1}\,\phi_i$. Thus ${\sf A}$ naturally defines a
complexified gauge field on $Q_3$. We have
\be {\sf F}^{2,0} = {\sf F}_{jk}dz^j \wedge dz^k  \ee
so the condition ${\sf F}^{2,0}=0$ simply becomes the condition that
the curvature of the complex connection ${\sf A}$ vanishes.
Decomposing in real and imaginary parts, we get the $F$-terms
\ba 0\ =\ {\rm Re}\, {\sf F}^{0,2} &=& F - [\phi_J,\phi_J] \eol 0\
=\ {\rm Im}\, {\sf F}^{0,2} &=& {D}_A\, \phi_J. \ea
Further, we have the $D$-terms
\be 0=g^{\mu\bar{\nu}} {\sf F}_{\mu\bar{\nu}} =  i\, K^{ij} D_i
\phi_{Jj} = i \,{D}_A^\dagger \phi_J \ee
These equations are precisely Hitchin's equations
\cite{HitchinHiggs,DonaldsonTwistedHarmonic}.

The equations for the Yang-Mills-Higgs fields on $Q_3$ are the
primary objects for our purposes. However if $Q_3$ admits an
integral affine structure, then the above isomorphism between
solutions of Hitchin's equations on $Q_3$ and the Hermitian
Yang-Mills equations on $TQ_3$ may be extended globally over $Q_3$.
In order to do this on $S^3$ we would need to excise a suitable
graph. Since the Higgs field takes values in the cotangent bundle
$T^*Q_3$, it will be useful to define dual coordinates:
\be d\tilde{y}_j = K_{jk} dy^k \ee
The K\"ahler form on $TQ_3$ naturally gets identified with the
standard symplectic form on $T^*Q_3$:
\be \omega = {i\over 2} K_{ij} dz^i \wedge d\bar{z}^j = K_{ij} dx^i
\wedge dy^j = dx^i \wedge d\tilde{y}_i \ee
For later use, we can also dualize $d\tilde{x}_j = K_{jk} dx^k$.
Then there is a natural complex structure also on $T^*Q_3$, in which
the complex coordinates are given by $\tilde z_j = \tilde x_j + i
\tilde y_j$, and a Calabi-Yau metric given by
\be ds^2 = K^{ij}(d\tilde{x}_id\tilde{x}_j +
d\tilde{y}_id\tilde{y}_j) \ee %
Furthermore, we will see in the next section that solutions of
Hitchin's equations can be interpreted as Lagrangian branes in
$T^*Q_3$. This mapping between ${\bf R}^3$-invariant solutions of
the Hermitian Yang-Mills equations and solutions of the
Yang-Mills-Higgs equations is a real version of the Fourier-Mukai
transform.

There are two basic type of solutions of the above Yang-Mills-Higgs
equations. Solutions with $[\phi,\phi]=0$ are said to be flat, and
with this restriction the above data describes a Higgs bundle. One
can also have solutions with $[\phi,\phi]\not = 0$. These will not
describe Higgs bundles or ALE fibrations, but rather (in the picture
described in section \ref{SpectralCover}) they will describe
coisotropic branes in $T^*Q_3$. Such configurations were proposed to
be relevant for moduli stabilization in \cite{Acharya:2002kv}. In
this paper we consider $[\phi,\phi]=0$.

We would like to elaborate a bit on the interpretation of the
$D$-terms. Let us say that ${\sf A}$ is an $Sl(n,{\bf C})$
connection. Given a solution to the $F$-term equations, in order to
write the $D$-term equation we need to split $\sf{A}$ up into a real
part $A$ and an imaginary part $\phi_J$. In general this cannot be
done canonically, but depends on a choice of Hermitian metric $h$,
which can be thought of as an equivariant map from the universal
cover
\be h:\,\widetilde{Q}_3 \  \to\ {\bf H}_n, \qquad {\bf H}_n =
Sl(n,{\bf C})/SU(n) \ee
Given such a metric $h$, the covariant derivative $D_{\sf{A}}$ will
generally not preserve it, but it can be split up as
\be D_{\sf A} =  D_A + \phi_J \ee
where $D_A$ preserves the metric. Furthermore $\phi$ is locally
identified with the derivative $\nabla h: T\widetilde{Q}_3 \to T{\bf
H}_n$, and $D_A$ is the pull-back of the Levi-Civita connection on
${\bf H}_n$. The $D$-term equation
\be D_A^\dagger \phi = D_A^\dagger \nabla h = 0 \ee
is precisely the requirement of harmonicity of $h$ as a map
$\widetilde{Q}_3 \to {\bf H}_n$. Hence the solution of the $D$-term
equation is also called the harmonic metric
\cite{DonaldsonTwistedHarmonic,CorletteHarmonic}.

Let us spell this out in a little bit more detail in the abelian
case. In this case ${\bf C}^*/U(1)= {\bf R}_{\geq 0}$ is the
one-dimensional version of hyperbolic space, $\nabla h = h^{-1} dh$
and $A = -\half h^{-1} dh$. The Hermitian metric can be written as
$h = e^f$ where $f: \widetilde{Q}_3 \to {\bf R}$. The $D$-term
equation says that
\be  D_A^\dagger \nabla h = d^\dagger df = 0 \ee
which says that $f$ is a harmonic function on $\widetilde{Q}_3$, or
that $\phi=df$ is a harmonic one-form on $Q_3$. Of course we can
also argue more directly that setting $A$ to zero locally and
solving $d\phi = 0\Rightarrow \phi = df$ implies that the $D$-term
can be written as $d^\dagger df = 0$. But the above point of view is
useful for establishing the existence of solutions in the
non-abelian case \cite{DonaldsonTwistedHarmonic,CorletteHarmonic}.

The $F$-term part of the Yang-Mills-Higgs equations are the critical
points of a Chern-Simons functional with complex gauge group, so we
identify this as the four-dimensional superpotential
\be\label{complexCS} W\ =\ {1\over 16\pi^2}\int_{Q_3} {\rm Tr}({\sf
A}\wedge d{\sf A} + {2\over 3} {\sf A}\wedge {\sf A} \wedge {\sf A})
\ee
In fact we could have gotten this more directly by applying the real
Fourier-Mukai transform to the holomorphic Chern-Simons functional
\cite{Leung:2000zv}. Further the $D$-term is the moment map for real
gauge transformations, with respect to the K\"ahler form associated
to the metric
\be g({\sf A},{\sf A})\ =\ \int_{Q_3} |{\sf A}|^2\ =\ \int_{Q_3} |A
+ i \phi|^2 \ee
Thus it can be obtained as the critical point of the $D$-term
potential:
\be V_D\  \sim\  \half \int | D_A^\dagger \phi|^2 \ee
when varied over
possible hermitian metrics $h$. These are the precise analogues of
the quantities we wrote earlier for $G_2$-manifolds in section
\ref{G2manifolds}, but they also incorporate non-perturbative states
that arise at ALE singularities.

\newsubsection{Spectral cover picture}

\subseclabel{SpectralCover}

 Suppose we are given a configuration for
the adjoints $\phi$ satisfying $[\phi,\phi]=0$. Then the three
components of $\phi$ may be diagonalized simultaneously, and we may
associate to $\phi$ its spectral data (i.e. its eigenvalues).
Conversely, the Higgs field $\phi$ may be reconstructed from its
spectral data. This picture yields $A$-model branes in an auxiliary
Calabi-Yau three-fold $X = T^*Q_3$. In this subsection we discuss
this construction in more detail. There is a completely analogous
construction in $F$-theory \cite{Donagi:2009ra,Hayashi:2009ge}
involving spectral covers in the canonical bundle over a complex
surface. For convenience we will take $Q_3 = S^3$.

In the case of $A_n$-fibrations, the spectral cover picture is more
than just an auxiliary construction, since it describes the
$D6$-branes that we see in the weak coupling IIa limit. This is not
a coincidence, because in the limit $vol(S^3) \to \infty$ the
worldvolume theory on the $D6$-branes is well approximated by the
maximally supersymmetric $7d$ YM theory.

Let us briefly recap the general structure. For the moment we will
restrict to $A_{n-1}$-fibrations, i.e. $7d$ gauge theory with gauge
group $G=SU(n)$. The twisted adjoint scalars of the $7d$ gauge
theory give a section of
\be \phi\ \in\ T^*S^3 \otimes Ad({\cal G}) \ee
where ${\cal G}$ is the principle bundle with gauge group $G$.
Consider the Hitchin map, which takes the Higgs field to the
coefficients of $s$ in
\be {\rm det}(sI - \phi) \in {\rm Sym}^n\,T^*S^3 \ee
Here $s$ is a local coordinate on the bundle direction of $T^*S^3$,
and ${\rm Sym}^n$ is the $n$th symmetric power. The eigenvalues
$\{\lambda_1, \ldots, \lambda_n\}$ of $\phi$ correspond to the zero
set of the above section. Each $\lambda_i$ is a one-form, so has
three-components. Thus the zero set defines an $n$-fold covering of
the zero section in $T^*S^3$. This is called the spectral cover for
the fundamental representation. We will denote it as $C_{(E,\phi)}$,
or simply $C_E$ for brevity, and the covering map by $p_C:C_E \to
S^3$. Since $\phi$ lives in the adjoint of $SU(n)$, the eigenvalues
add to zero on each fiber of $T^*S^3$:
\be \lambda_1 + \ldots+ \lambda_n = 0 \ee
The gauge field $A$ gives a flat connection on a bundle $E$
associated to the fundamental representation of $SU(n)$. We can
think of $\phi$ as a map
\be \phi:E \to E \otimes T^*S^3 \ee
Then on each fiber $E$ decomposes into a sum of eigenspaces
$\oplus_i {\bf C} \ket{i}$ under the action of $\phi$. The
assignment $\lambda_i \to {\bf C}\ket{i}$ gives a line bundle $L_E$
on $C$ called the spectral line bundle, and since $D_A \phi = 0$,
$D_A$ commutes with the action of $\phi$ on $E$ and therefore $A$
gives a flat connection on this line bundle. Conversely, given a
spectral cover $C_E$ together with a flat line bundle $L_E$, we can
reconstruct the Higgs bundle on $S^3$ by
\be E=p_{C*} L_E, \qquad \phi = p_{C*}s \ee
which yields a rank $n$ bundle $E$ and a map $\phi:E \to E \otimes
T^*S^3$. In order for this to be an $SU(n)$ bundle, rather than a
$U(n)$ bundle, we must have
\be {\rm det}(p_{C*}L_E) = 1 \ee
where $1$ denotes the trivial line bundle. This puts a topological
constraint on the allowed line bundles on the spectral cover.

We claim that the spectral cover yields an $A$-type brane in
$T^*S^3$. In order to see this, we may analyze the $F$- and
$D$-terms locally. Since the gauge field $A$ is flat, it may locally
be gauged away. Further, as we discussed the equation $D\phi = 0$
splits up into $n-1$ abelian equations $d\phi = 0$, where in the
last equation we used $\phi$ to denote an abelian Higgs field. It is
well known that the condition that $\phi$ be closed is equivalent to
the section being Lagrangian with respect to the standard symplectic
form on the cotangent bundle. To see this, writing $\phi = \phi_i
dx^i$, the equation $\omega|_C=0$ gives
\be 0 = dx^i \wedge d\tilde y_i|_C = dx^i \wedge d\phi_i(x) =
-d(\phi_i dx^i)= -d\phi \ee
Thus the $F$-terms equations correspond precisely to the condition
that the spectral cover is a Lagrangian submanifold of $T^*S^3$,
together with a flat connection. It is also worth noting that $d\phi
= 0$ implies that $\phi = df$ for some function $f$ on a local patch
of $S^3$. Thus each sheet of the spectral cover may locally be
represented as the graph of the differential of a real-valued
function.

Naively one might expect these branes to be special Lagrangian also,
since this is the usual requirement for supersymmetric $D6$ branes
in IIa string theory. However, this is not quite the case. Locally
on each sheet we have
\be \Omega^{3,0}|_C = d\tilde z_1\wedge d\tilde z_2\wedge d\tilde
z_3|_C = \det(I + i{\rm Hess}(f)) d\tilde x_1 \wedge d\tilde x_2
\wedge d\tilde x_3 \ee
where we used $d\tilde z_i = d\tilde x_i + i d\tilde y_i$ and $
\tilde y_i(x) dx^i = \phi=df$. Requiring that the imaginary part
vanishes identically leads to the non-linear equation
\be\label{MOLaplace} \Delta f = \det({\rm Hess}(f)) \ee
It's not hard to see where this apparent discrepancy comes from. The
right hand side comes from a higher derivative correction to the
two-derivative SYM theory. Indeed one may write a similar non-linear
correction term for $10d$ SYM theory, which is related to
(\ref{MOLaplace}) by the Fourier-Mukai transform\footnote{This is
for the abelian case; the non-abelian case is apparently not yet
completely understood}\cite{Leung:2000zv}:
\be \omega \wedge \omega \wedge F = F \wedge F \wedge F \ee
In the large volume limit in which we are working, such higher
derivative corrections are parametrically small and can be neglected
to first approximation. Thus in our approximation, $D6$-branes in
type IIa and more generally spectral covers of ALE fibrations/Higgs
bundles are described by `harmonic' Lagrangian branes in $T^*S^3$,
rather than special Lagrangian branes.

In the IIa context it is natural to conjecture that the `harmonic'
Lagrangian flows to a unique special Lagrangian under mean curvature
flow \cite{Thomas:2001vf}. In the $M$-theory context however it
seems inappropriate to look at special Lagrangians. The reason is
that in the IIa context we have two expansion parameters in $4d$,
namely $g_s$ and $\ell_s/R_{KK}$, but in the present context there
is only a single expansion parameter (namely $1/vol(S^3)$ in Planck
units), and so the higher order corrections discussed above can
compete at the same order in this expansion parameter with other
corrections such as KK loops. It would be physically incorrect to
include only one type of correction and ignore the other
contributions at the same order in the expansion parameter.

\begin{figure}[t]
\begin{center}
            \scalebox{.45}{
               \includegraphics[width=\textwidth]{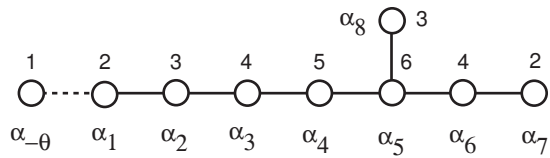}
               }
\end{center}
\vspace{-.5cm} \caption{ \it The extended $E_8$ Dynkin diagram and
Dynkin indices.}\label{E8Dynkin}
\end{figure}

For applications to $M$-theory phenomenology we are interested in
$E_8$ Higgs bundles. The spectral cover in this case can get quite
complicated, but fortunately phenomenological considerations dictate
that we only consider non-trivial configurations for an $SU(n)$
subbundle (so that the Higgs field breaks the $E_8$ gauge group to
the commutant of $SU(n)$; the unbroken part of the gauge group is
called the GUT group). In particular we would like to consider the
case $n=5$, which yields an $SU(5)$ GUT group. The $E_8$ spectral
cover has 248 sheets and decomposes into several pieces, according
to the decomposition\footnote{The cover corresponding to $({\bf
24},1)$ further splits into two pieces, a cover of degree 20 and a
four-fold multiple of the zero section.}
\be {\bf 248} = ({\bf 24},{\bf 1}) + ({\bf 1},{\bf 24}) + (\bfv,\bt)
+ (\bfb,\btb) + (\bt,\bfb) + (\btb,\bfv)  \ee
of $E_8$ under $SU(5)_{GUT} \times SU(5)_H$. The most important is
the spectral cover for the fundamental representation of $SU(5)_H$,
which determines all the others uniquely. This cover intersects each
fiber of $T^*S^3$ in five points $\lambda_1, \ldots,  \lambda_5$,
with
\be \lambda_1 + \ldots + \lambda_5 = 0 \ee
Here addition is defined in the obvious way in each fiber. In the
language of ALE fibrations, the $\lambda_i$'s correspond to certain
$FI$ parameters of the $E_8$ ALE using the dictionary described in
section \ref{ALEfibrations}. Using the labelling in figure
\ref{E8Dynkin}, we may take the $\lambda_i$ to describe the size
parameters of the following exceptional cycles:
\be
\begin{array}{rclrcl}
 \ket{1} &=& \alpha_4  & \ket{4} &=& \alpha_1 +\alpha_2 + \alpha_3 + \alpha_4  \eol
\ket{2} &=& \alpha_3 +\alpha_4 & \ket{5} &=& \alpha_{-\theta} +
\alpha_1 + \alpha_2 + \alpha_3 + \alpha_4 \eol \ket{3} &=& \alpha_2
+ \alpha_3 + \alpha_4 \qquad  & & &
\end{array}
 \ee
The sizes of the cycles $\{\alpha_5, \ldots ,\alpha_8\}$ are taken
to be zero, generating an $SU(5)$ GUT group, and all other cycles
are obtained as linear combinations.
As we will discuss in more detail later, when one of the
$\lambda_i$'s goes to zero, i.e. when the cover intersects the zero
section, one may get a chiral or anti-chiral field in the $\bt$
localized here. Another piece of the $E_8$ cover corresponds to the
anti-symmetric representation of $SU(5)_H$. The cover
$C_{\Lambda^2E}$ intersects each fiber in
\be \lambda_i + \lambda_j, \qquad i < j. \ee
and we will see later that when this cover intersects the zero
section, we may get a chiral or anti-chiral field in the $\bfb$
localized there.

To summarize, we've gone through the following chain:
\be
\begin{array}{c}
  \frame{\makebox(150,25){$G_2\ {\rm metric}$ and flat $C_3$}} \\[3mm]
  \updownarrow \\[4mm]
  \frame{\makebox(150,25){ $d(C_3 + i\Phi) = d^*\Phi = 0$}} \\[3mm]
  \wr \\[3mm]
  \frame{\makebox(150,25){ { Higgs\  bundle} }}\\[3mm]
  \updownarrow \\[3mm]
  \frame{\makebox(150,25){ {
h-Lag\ branes\ in\ }$ T^*Q_3$}}
\end{array}
  \ee
The conjecture is that we can also go in the reverse, i.e. given
solutions to Hitchin's equations with $[\phi,\phi]=0$ one may
reconstruct solutions to $ d\Phi = d^*\Phi = 0$. This must be
correct if the $7d$ gauge theory is to give an accurate description
of $M$-theory dynamics on an ALE singularity, as is expected
physically for large $S^3$.

\newsubsection{Non-compact branes}

We claim that in order to get interesting solutions, we have to
allow for certain source terms in the YMH equations. To see this,
let us assume that we do not have any source terms. Now flat
connections are characterized by their monodromies, and since
$\pi_1(S^3) = 1$ any flat connection is equivalent to the trivial
connection. This is in accord with the statement that branes wrapped
on the minimal three-sphere in $T^*S^3$ do not form bound states
\cite{Strominger:1995cz}.

One may get non-trivial solutions by instead quotienting the $S^3$
by a freely acting discrete group $\Gamma$, so that
$\pi_1(S^3/\Gamma)$ will be non-trivial. However the non-trivial
bound states one can make are still not of the type we want. We need
the discrete group $\Gamma$ to be freely acting, and in this case
must be finite (basically $\Gamma$ is a product of ADE subgroups of
$SU(2)_L \times SU(2)_R$ acting on $S^3$). Therefore the monodromies
must be contained in the compact part of the complexified gauge
group. This means that Wilson lines for the gauge field can be
turned on but the Higgs field has zero expectation value. In order
to get more interesting solutions with a non-trivial Higgs field we
need to do something else.

In order to get interesting solutions we need to allow for
non-compact flavour branes, i.e. source terms in the
Yang-Mills-Higgs equations. This is completely analogous to the
meromorphic Higgs bundles appearing in local $F$-theory models
\cite{Donagi:2009ra,Hayashi:2009ge}. One may also see how this
arises by applying heterotic/$M$-theory duality to heterotic models.
Let us take for instance a heterotic model with the spin connection
embedded in the gauge connection. In a suitable limit, the heterotic
Calabi-Yau three-fold admits a Lagrangian $T^3$-fibration over
$S^3$.
The Wilson lines of the bundle along the $T^3$ fibers form a
covering of the $S^3$, which consist of three-points on the generic
dual $T^3$ fiber. This is the heterotic picture of the spectral
cover that we discussed, for the special case of the tangent bundle.
However over special subsets in $S^3$ the cover may wrap some of the
circles of the $T^3$-fibration. Eg. over a graph in $S^3$ it may
contain a $T^2 \subset T^3$, and over special points in $S^3$ it may
contain the whole $T^3$.
As in $F$-theory/heterotic duality, we expect that in a suitable
limit this data is equivalent at the level of $F$-terms to a
degenerate $K3$ fibration over $S^3$ with a section of $E_6$ $ALE$
singularities, although for $M$-theory/heterotic duality this is of
course not established. Taking the local limit, we should take the
size of the dual $T^3$ fibers to infinity in the heterotic picture.
This gives the picture advocated above of a Higgs field which is
generically finite, but may blow up over a special subset of $S^3$.

\newpage

\newsection{The effective theory}

In the previous section, we reformulated the problem of constructing
phenomenological $M$-theory compactifications in terms of Higgs
bundles and spectral covers. In this section we explain how the low
energy degrees of freedom and their interactions arise from the
compactified Yang-Mills theory. Qualitatively this is already
largely understood in the literature, but in order to construct
models and carry out the computations explicitly we need some new
tools. Therefore we will reformulate some old results in our present
language and introduce Morse theory in order to relate the spectrum
to more readily computable quantities.

\newsubsection{Chiral matter}

\subseclabel{ChiralSpectrum}

Intuitively, chiral matter will be localized on some kind of
solitonic configuration of the Higgs field. At the center of such a
soliton, one of the eigenvalues of the Higgs field is going to zero.
Thus we would like to analyze the Dirac equation in such a solitonic
background. In non-degenerate situations, there is only a single
eigen-value or combination of eigen-values of the Higgs field going
to zero. Therefore in the non-degenerate case it is sufficient to
consider abelian Higgs fields only, and we will assume this through
much of the discussion below. As we will discuss in the next
subsection however, there are some global effects which may require
us to look at the full non-abelian field.

Recall that locally we can set the gauge field to zero, and write
$\phi = df$. Therefore the zeroes of $\phi$ are sometimes also
called `critical points'. We will now describe how this gives rise
to chiral matter. This is essentially already discussed in the
literature, particularly \cite{Berkooz:1996km,Acharya:2001gy} for
the present setting, but will be reformulated somewhat to fit our
purposes.

In order to calculate the spectrum, we need to solve a Dirac
equation with a background Higgs field turned on. Following
\cite{Witten:1982im} we will rescale the Higgs field by a positive
real number $t\sim 1/\hbar$ and calculate the spectrum in a $1/t$
expansion.

It is convenient to think of the spinors in the $7d$ Yang-Mills
theory from a ten-dimensional point of view. The ten-dimensional
${\bf 16}$ decomposes under $SO(3,1) \times SO(6)$ as
\be\label{16decomposition} ({\bf 2,1,4}) + ({\bf 1,2,\bar{4}}) \ee
which are further related by the Majorana condition. The ${\bf 4}$
and $\bar{\bf 4}$ of $SO(6)$ may be identified with $(0,p)$ forms,
with $p$ even odd for ${\bf 4}$ and $p$ even for $\bar {\bf 4}$.
Each breaks up under $SO(3)_Q \times SO(3)_R$ as a $({\bf 2,2})$,
but they have different eigen-values under $\Gamma_6$. We can denote
them as $({\bf 2,2})_\pm$ according to their $\Gamma_6$ eigenvalue.

As we discussed, due to twisting needed to maintain $N=1$
supersymmetry, we can identify $SO(3)_R$ with $SO(3)_{Q_3}$. This
diagonal $SO(3)$ may be identified with the real subgroup $SO(3)
\subset SU(3)$ of the Calabi-Yau holonomy group fixed by an
anti-holomorphic involution, which gives another way to see the
twisting. In any case, our spinors are functions of $x_i$ and
transform as spinor bilinears on $Q_3 = S^3$. As is well known, such
bilinears can be identified with differential forms on $Q_3$, i.e.
they may be identified with sections of
\be\label{twistedspinors} \psi\ \in\ {\cal A}^p(Q_3, {\bf C})
\otimes Ad({\cal G}) , \qquad p=0,1\ee
where ${\cal G}$ is the principal bundle with gauge group $G$. Note
that this matches with the bosonic field content. In order to relate
this to the description above, given a $p$-form wave-function $\psi$
we may associate with it a $(0,p)$-form by replacing $dx^i \to d\bar
z^i$ in $\psi$, and and a $(p,0)$-form by replacing $dx^i \to
d{z}^i$ in $\psi^*$. By Serre duality (i.e. taking the complex
conjugate and contracting with $\Omega^{3,0}$) we may relate the
$(p,0)$-forms to $(0,3-p)$-forms which transform in the same
representation. The reason for replacing $\psi^*$ by its Serre dual
is that the Dirac operator acts more naturally in this basis. In
terms of the original $p$-form $\psi$ this is just the real Hodge
$*$-operator $\psi \to
*\psi$ on $Q_3$ without complex conjugation. So it is natural to
allow differential forms (\ref{twistedspinors}) for all values of
$p$, with equivalent degrees of freedom related by the $*$-operator.
Tracing back to (\ref{16decomposition}) we see that chiral fermions
are naturally paired with odd $p$-forms and anti-chiral fermions are
paired with even $p$-forms.

We assume that we have a gauge group $G$ which is broken to a
subgroup $H$ by turning on an abelian component of the Higgs field.
We decompose the adjoint representation of $G$ under $H \times U(1)$
as
\be Ad(G)\ =\ Ad(H)_0 + R_q(H) + \overline{ R}_{-q}(H) + {\bf 1}_0
\ee
The worldvolume gauge fields can be set to zero locally. The Dirac
operator acting on the spinors in the $R$-representation is then
given by
\be i{\sf Q}_t\ =\ i\, \sum_{j=1,2,3}(\del_j + t\del_j
f)(a_j^\dagger + a^j) \ee
Here we set $q\to 1$ because its precise value is inconsequential,
only its sign is important. The Dirac operator acting on spinors in
the $\bar R$-representation is given by interchanging $f \to -f$.
When identifying the spinors with forms, the Clifford algebra may be
represented as $a_i^\dagger = \wedge dx_i$ and $a^i = \imath_{\del
/\del x_i}$, so we get
\be {\sf Q}_t\ =\ d_t +d_t^\dagger \ee
where
\be d_t\  =\ d + t\,df \wedge \ee
The operator ${\sf Q}_t$ is exactly the operator discussed at length
in \cite{Witten:1982im}, so we will borrow from the discussion
there.

The operator ${\sf Q}_t$ may be thought of as a supercharge for
supersymmetric quantum mechanics with target space $M$. The
hamiltonian of this system is given by
\be {\sf Q}_t^2 = H_t = \Delta + t^2 (df)^2 + \sum_{i,j} t
{D^2f\over Dx_i Dx_j} [a^{\dagger i},a^j] \ee
where $\Delta$ denotes the usual Laplace-Beltrami operator. For
large $t$, the Hamiltonian is dominated by the potential energy
$|df|^2$. In order to minimize the potential energy for large $t$,
the eigenfunctions must be peaked around the critical points of $f$.
Therefore we may focus on a single critical point and approximate
$f$ by a quadratic potential. Up to coordinate transformations, $f$
may locally be written as:
\be\label{localfform} f= \half \sum_{i=1,2,3} p_i x_i^2 \ee
The $p_i, i=1,2,3$ are real constants. They are all non-zero because
we assumed that the critical points are non-degenerate. Then the
Higgs field near the critical point is given by
\be \phi = df = \sum_{i=1,2,3} p_i x_i\, dx_i \ee
The $D$-term equation put a restriction on the $p_i$:
\be {\rm Tr \ Hess}(f) = p_1 + p_2 + p_3 = 0 \ee
%

 \begin{figure}[t]
\begin{center}
            \scalebox{.6}{
               \includegraphics[width=\textwidth]{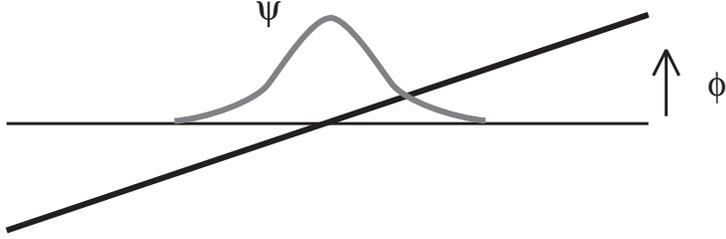}
             }
\end{center}
\vspace{-.5cm}
\begin{center}
\parbox{14cm}{\caption{ \it Fermion zero mode localized on a defect.}\label{ChiralFermion}}
\end{center}
 \end{figure}
Using coordinates in which $f$ takes the diagonal form
(\ref{localfform}), we may clearly use separation of variables. Thus
we concentrate on one variable $x_i$ temporarily and finally tensor
the wave functions together. Then we have a standard domain wall
set-up (see Figure \ref{ChiralFermion}). The Dirac equation becomes
\be  \left[{\del\over \del {x_1}}+ t p_1 x_1\right] \psi_1^+ = 0,
\qquad \left[{\del\over \del {x_1}} - t p_1 x_1\right] \psi_1^- = 0
\ee
The local solution is
\be \psi_1^\pm(x) \ =\ e^{\pm t p_1\, x_1^2/2} \epsilon^\pm \ee
where $\epsilon^+ = 1$ and $\epsilon^- = dx_1$. Inspecting the
exponential factor, we see that either $\psi_1^+$ is normalizable
and $\psi_1^-$ is not, or vice versa. The normalizable solution is
physically sensible and the non-normalizable one should be
discarded. Which solution is normalizable clearly depends only on
the sign of $p_1$.

Tensoring together with $\psi_2$ and $\psi_3$ and a four-dimensional
chiral or anti-chiral spinor $\chi^\pm$, we get the full wave
function:
\be   \chi^\pm \psi_1^\pm \psi_2^\pm \psi_3^\pm \otimes R({\cal H})\
\in\ {\cal A}^*(Q_3;{\bf C})\otimes R({\cal H}) \ee
As we discussed, not all of these combinations are allowed. The four
dimensional chirality is correlated with the degree of the form,
which now becomes the number of negative eigenvalues. The number of
negative eigenvalues of the Hessian at an isolated critical point is
called the Morse index. The cases of Morse index zero or three (i.e.
$(+,+,+)$ and $(-,-,-)$) are ruled out by the $D$-terms, since a
harmonic function can not have local minima or maxima. Therefore up
to permutations, we have Morse index one $(+,+,-)$ which gives a
chiral fermion, and Morse index two $(+,-,-)$ which gives an
anti-chiral fermion.

In addition, we should analyze the Dirac equation for the remaining
pieces in the decomposition of $Ad(G)$, namely $Ad(H)_0$ and $1_0$.
The corresponding zero modes are not localized at the critical
points and we need some global information. In this case the Dirac
operator is just given by the exterior differential, and the zero
modes are in one-to-one correspondence with Betti numbers. From
$Ad(H)_0$ we get a gaugino transforming in $Ad(H)$, and $b_1(Q)$
adjoint chirals. From the $1_0$ we get $H^1(C)$ moduli (where $C$ is
the spectral cover). This concludes the derivation of the massless
spectrum to all orders in the $1/t$ expansion.

As a simple example, consider the local unfolding of an $SU(6)$
singularity to an $SU(5)$ singularity, considered in
\cite{Acharya:2001gy,Atiyah:2001qf}. That is, we will consider an
$A_5$ ALE surface fibered over ${\bf R}^3$, such that for $\vec{x}
=\vec{0}$ all the vanishing cycles are zero size and we have an
$A_5$ singularity, and such that for $\vec{x } \not = 0$ the ALE is
partially resolved, but we still have an $A_4$ singularity. Then we
need to turn on an abelian Higgs field. Under $SU(5)\times U(1)_Q$,
the adjoint of $SU(6)$ decomposes as
\be Ad(SU(6)) = Ad(SU(5)) + {\bf 5} + \bar{\bf 5} + {\bf 1} \ee
and so we expect a chiral fermion at $\vec{x}=0$ which transforms as
a $\bfv$ or $\bfb$ under the unbroken $SU(5)$ gauge symmetry. Let us
first phrase the configuration in our current language, and then
compare with \cite{Acharya:2001gy}.

 \begin{figure}[t]
\begin{center}
            \scalebox{.45}{
               \includegraphics[width=\textwidth]{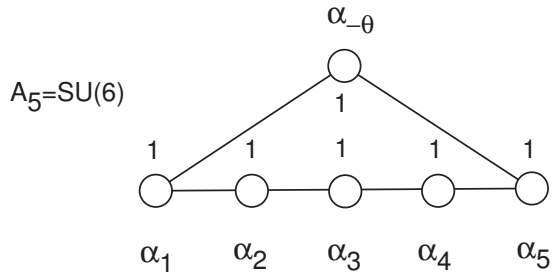}
               }
\end{center}
\vspace{-.5cm} \caption{ \it The extended $A_5$ Dynkin diagram and
Dynkin indices.}\label{A5Dynkin}
\end{figure} 

Our Higgs field will be proportional to a Cartan generator $U(1)_Q$
which is embedded in $SU(6)$ as $\omega_Q ={\rm
diag}(1,1,1,1,1,-5)$. In terms of the canonical basis $\omega_k$
satisfying
\be [\omega_k, \alpha_l] = \delta_{kl} \ee
this corresponds to $\omega_Q =6\omega_5$. To each node we can
associate an abelian Higgs field $\phi_k \omega_k$, whose three
components are the three FI parameters for the corresponding cycle
$\alpha_k$. They satisfy the constraint
\be d_{-\theta} \phi_{-\theta} + \ldots + d_5 \phi_{\alpha_5} = 0
\ee
where $d_k$ are the Dynkin indices. In the present case, the $d_k$
are all equal to one. (This description is redundant because we can
always use this relation to eliminate $\phi_{-\theta}$, but it
becomes quite convenient for non-abelian Higgs field VEVs). Now we
set
\be {\phi}_{\alpha_5} = -{ \phi}_{-\theta} = df, \qquad f \propto
\sum_{i=1}^3 p_i x_i^2 \ee
Since $\{\phi_{\alpha_1}, ..,\phi_{\alpha_4}\}$ are all kept zero,
the corresponding cycles $\alpha_k$ are all kept at zero size, and
an $SU(5)$ singularity is preserved. This satisfies the $F$-terms,
and using the metric $ds^2 = \sum dx_i^2$ on ${\bf R}^3$ it also
satisfies the $D$-terms provided $p_1 + p_2 + p_3$ = 0. By our
previous analysis, we get a chiral fermion localized at
$\vec{x}=\vec{0}$, in the $\bfv$ if $p_1 p_2 p_3>0$ or in the $\bfb$
if $p_1 p_2 p_3<0$.

This description agrees with the hyperk\"ahler quotient construction
of \cite{Acharya:2001gy}. Their $D$-terms are given by $\vec{D} =
(a, \bar{b})/U(1)$, where the $U(1)$ acts with charge one on $a$ and
$b$. The authors of \cite{Acharya:2001gy} choose the unfolding
$\vec{x} = (a,b)/U(1)$, i.e. we changed the sign of one of the
components of $\vec{D}$ (changing the Morse index from zero to one)
and then identified the image with ${\bf R}^3$. Thus this agrees
with our claims above except for an inconsequential rescaling in the
metric on ${\bf R}^3$. Additional constructions along the lines of
\cite{Acharya:2001gy} can be found in
\cite{Berglund:2002hw,Bourjaily:2009vf}.

Of course we want a gap between the massless modes and the KK modes
of the $7d$ gauge theory, so we are not interested in fibering over
${\bf R}^3$ but over a compact three-manifold like $S^3$ or
$S^3/\Gamma$. Then there will necessarily be higher order terms in
$f$ and additional critical points. Our calculation was not exact,
and in fact there are corrections to $H_t$ exponentially suppressed
in $t$ which may lift some of the zero modes we found. These
corrections are the topic of section \ref{MorseCohomology}.

We can also state the formula for chiral matter in terms of the
spectral cover in the auxiliary Calabi-Yau $T^*S^3$. Locally we can
describe a sheet by the graph of $df$. A critical point of $f$
corresponds to an intersection point between the graph of $df$ and
the zero section, and the sign of the intersection is just $(-1)^p$
where $p$ is the Morse index. Thus the statement is that one must
count with sign the number of intersection points of a harmonic
Lagrangian brane $C$ with the zero section $C_0$:
\be N_\chi(R) = \# (C_0 \cap C)_-, \qquad N_{\chi}(\bar R) = \# (C_0
\cap C)_+ \ee
Consider for instance $E_8$ models broken to $SU(5)$ by an $SU(5)$
Higgs field. The $SU(5)$ Higgs field can be encoded in a Lagrangian
brane $C_E$ in $T^*S^3$ which is a five-fold covering of the zero
section, or in a ten-fold covering $C_{\Lambda^2E}$ associated to
the anti-symmetric representation.  When $C_E$ intersects the zero
section, we have $\lambda_i \to 0$ for some $i$ and the symmetry is
locally enhanced to $SO(10)$.  Using the decomposition
\be Ad(SO(10)) = Ad(SU(5)) + \bt + \btb + {\bf 1} \ee
the chirals in the $\bt$ are counted by the number of negative
intersections between $C_E$ and the zero section. Similarly when
$C_{\Lambda^2E}$ intersects the zero section, $\lambda_i +
\lambda_j\to 0$ for some $i\not = j$ and the symmetry is locally
enhanced to $SU(6)$, i.e. it locally looks like the example
discussed above. Thus the number of $\bfv$ or $\bfb$'s is counted by
the intersection points of the cover $C_{\Lambda^2E}$ with the zero
section. However pairs of chirals and anti-chirals may still be
lifted through instanton effects, leading to the quantum
intersection theory of Lagrangian branes in $T^*S^3$.

\newsubsection{Abelian solutions}

Recall that in order to get non-trivial solutions to Hitchin's
equations on $S^3$ or $S^3/\Gamma$, we have to allow for non-compact
branes, i.e. we have to allow for singularities in the Higgs field.
We will generally assume that these singularities are located on a
graph $\Delta$ in $S^3$, although one could consider more general
situations. Then locally we may choose a coordinate $r$ transverse
to the graph, and an angle $\theta$ in the plane transverse to the
graph. In the abelian case, the local behaviour as $r \to 0$ is
\ba A&\sim &\alpha d\theta \eol \phi &\sim& \beta d\log r + \gamma
d\theta \ea
The parameters $\alpha, \beta$ and $\gamma$ are boundary data that
we have to specify. In this section we discuss the special case when
there is no monodromy of the gauge and Higgs fields, i.e. $\alpha +
i \gamma = 0$, and we return to the general case later.

We excise a small tubular neighbourhood of $\Delta$ from $S^3$,
which therefore becomes a manifold-with-boundary which we will
denote by $M$. Then the $F$-terms on $M$ simply read
\be d\phi = 0 \ee
and since the Higgs field carries no monodromy ($\gamma = 0$), we
may express $\phi = df$ for a globally defined function $f$. The
$D$-terms read
\be d^* df = 0 \ee
on the complement of $\Delta$. In fact it is simpler in this case to
think of $\beta$ as a charge density along $\Delta$, and write the
$D$-terms as a Poisson equation on $S^3$
\be d^* df = \beta \ee

The above equations are of course very familiar. We are simply
dealing with an electro-statics problem, with the Higgs field $\phi$
playing the role of the (dual of the) electric field, $\beta$
playing the role of a charge density along the graph, and $f = \log
h$ playing the role of the electro-static potential. Thus we can
solve this problem in the standard way, by using the Green's
function for the Laplacian. There is a single consistency constraint
that needs to be satisfied: using the divergence theorem, we get
\be 0 = \int_M d^* \phi = \int_{\del M} \phi \cdot \nu \ee
and hence the total flux through the boundary must vanish, which
means that the total `charge' on $S^3$ must vanish (Gauss's law).
Accordingly there must be positive and negative charges, and we can
split up the graph $\Delta$ as $\Delta^+ \cup \Delta^-$ which are
positively and negatively charged respectively. Let us denote the
Green's function for the Laplacian as $G(x,y)$. The solution to the
BPS equations is given by
\be \phi = df, \qquad f(y) = \int_{S^3}d^3 x \sqrt{g}\, \beta(x)
G(x,y) \ee

\newsubsection{Instanton corrections and Morse cohomology}
\subseclabel{MorseCohomology}

In the above we saw that one may construct abelian solutions by
solving a simple electro-statics problem. However even though we
know there exists a solution satisfying both the $F$-terms and the
$D$-terms, one can learn much by imposing only the $F$-term
equations. That is, in the following we will assume that $\phi$ is
closed but not necessarily harmonic. Recall that the spectrum of
massless charged chiral and anti-chiral matter is related to the
critical points of $\phi$. As we will now discuss, the massless
spectrum may actually be deduced using purely combinatorial methods
and is common to all solutions of the $F$-terms, independent of
whether $\phi$ is harmonic.

Let us denote the set of critical points of index $i$ by ${\rm
Crit}_i(f)$. It is a well-known fact that on a compact manifold the
number of critical points of Morse index $i$ is bounded below by the
$i$-th Betti number:
\be \#\,{\rm Crit}_i(f)\ \geq\ h^i(M) \ee
This is the weak form of the Morse inequalities, one of the central
results of Morse theory (a brief review may be found in appendix A).
In the present setting we also have boundaries, and we really get a
bound in terms of betti numbers for relative cohomology. For
simplicity let us temporarily ignore this issue. The point is that
we get a lower bound on the massless spectrum in terms of
topological data.

Now it turns out the situation is actually much better than that.
The calculation of the massless spectrum in section
\ref{ChiralSpectrum} was exact to all orders in a $1/t$ expansion,
but may still be corrected by instantons. Once we take these quantum
effects into account, we will find that the massless spectrum is in
fact {\it exactly} computed by the Betti numbers. In other words, we
may read of the massless spectrum just from the topology of $M$,
which may be computed by purely combinatorial methods. Following
\cite{Witten:1982im} (see also \cite{Salomonson:1981ug}), we will
now briefly explain how instantons correct the computation.

The potential energy function $V \sim |df|^2$ has multiple critical
points. However it is not generally true that the states we found at
each critical point are all true ground states. We haven't yet
accounted for the possibility of tunnelling. To see this, we
consider two critical points $p,q$ and their associated ground
states $\ket{p},\ket{q}$ and compute the amplitude $\vev{q|Q_t p}$.
The effective Lagrangian of the $7d$ SYM describing excitations
along $M$ in the representation $R$ is that of supersymmetric
quantum mechanics on $M$:
\be \mathscr{L} = \half (dx)^2 -\half t^2(df)^2 + {i\over 2}\bar
\psi (D + t D^2f)\psi +{1\over 4} R \bar \psi \psi \bar \psi \psi
\ee
The fermions may be thought of as the operators $a^i \sim \wedge
dx^i$ encountered earlier. Ignoring the fermions, the Euclidean
action is given by:
\ba 2\,S_E &=& \int d\lambda \, (dx)^2 + (t df)^2 = \int d\lambda(dx
\pm t df)^2 \mp t\int d\lambda df \eol &\geq& t\,|f(q) - f(p)| \ea
where $\lambda$ denotes Euclidean time. The potential is then turned
upside down, and there are instanton solutions:
\be {\del x_i\over d\lambda} = \pm t g^{ij}{\del f\over \del x_i}
\ee
with $ x(\lambda\to -\infty) = q, \ x(\lambda\to +\infty) = p$.
These are gradient flow trajectories that connect critical points
$p$ and $q$. For instanton contributions to $d_t$ we want the minus
sign above. Each such instanton contributes
\be \vev{q|d_t| p} \sim  \int dx_0 d\bar\psi_0 {\det_F\over
\det_B^{1/2}}\ d_t|_{class} e^{-\half t(f(q)-f(p))} \ee
Here $x_0$ and $\bar \psi_0$ denote the bosonic and fermionic zero
modes of the instanton; $\det_B$ and $\det_F$ denote the bosonic and
fermionic fluctuation determinants; and we evaluate $d_t$ on the
instanton solution.

To get the amplitude, we need to compute the one loop fluctuation
determinant. In supersymmetric theories, the determinants for the
non-zero modes of the bosons and fermions cancel, but there are zero
modes. For an instanton connecting two critical points whose Morse
index differs by one, there is one bosonic zero mode (for the broken
translation invariance in Euclidean time) and one fermionic zero
mode (for the broken supersymmetry generator). More generally, the
number zero modes is given by the Morse index of $p$ minus the Morse
index of $q$. Since the operator $d_t = \bar\psi(\del x+ \del f)$
soaks up exactly one bosonic and one fermionic zero mode, the
instanton only contributes if the difference in the Morse indices is
equal to one.

We further need to know the coefficient, in particular the sign of
the coefficient. This is slightly subtle since the coefficient is
proportional to $1/t$, so we simply state the result. Consider the
subspace $V_p$ of the tangent space at $p$ on which the Hessian is
negative definite. Its dimension is given by the Morse index of $p$
and it carries an orientation induced by the differential form
$\ket{p}$. Let $\vec{v}$ denote a vector tangent to the gradient
flow trajectory, and $V_p^\perp$ the subspace of $V_p$ orthogonal to
$\vec{v}$. It carries an orientation induced by the contraction of
$\ket{p}$ with $\vec{v}$. At $q$ we have an analogous subspace $V_q$
with orientation induced by $\ket{q}$. Now we transport $V_p^\perp$
along the gradient flow trajectory to $q$. If the orientation agrees
with $V_q$ we use the plus sign, and if they disagree we use the
minus sign.

Hence we deduce that non-perturbatively in $1/t$, the action of
$d_t$ on the would-be zero modes has the following correction:
\be d_t \ket{p} = \sum_{C^{i+1}(M)} n(p,q) e^{-\half t(f(q)-f(p))}
\,\ket{ q} \ee
where $n(p,q)$ counts with sign the number of trajectories. By
rescaling $\ket{p}\to e^{\half tf(p)}\ket{p}$, $\ket{q}\to e^{\half
tf(q)}\ket{q}$ we can get rid of the exponential factors, and hence
we simply get
\be d_t \ket{p} = \sum_{C^{i+1}(M)} n(p,q) \,\ket{ q} \ee
Since $d_t^2 = 0$, it is a boundary operator. The actual zero modes
are those which are also annihilated by $d_t $ and $d_t^\dagger$, so
the true zero modes are in one-one correspondence with the
cohomology of $d_t$. This is the Morse complex.

 It's not hard to recover these instantons in
the other pictures we have been using. In the ALE fibration picture,
the gradient flow trajectories connecting critical points can be
interpreted as membrane instantons. Let's assume assume for instance
that we have an $A_1$ ALE space fibered over $S^3$, and our line
bundle is embedded as the diagonal generator in $SU(2)$, generically
breaking the $SU(2) \to U(1)$. Then in the ALE sitting over the
critical point of $f$, an $S^2$ shrinks to zero size and the $U(1)$
gets enhanced to $SU(2)$ due to a massless $M2$ brane wrapping the
$S^2$. Now let us take this $S^2$ and transport it along the
gradient flow trajectory to the other critical point. In this way we
trace out an $S^3$ inside the ALE fibration which projects to the
gradient flow trajectory in the base $Q_3$, and there is a membrane
instanton obtained by wrapping an $M2$-brane on this $S^3$. The
action of this instanton is proportional to the area of the $S^3$,
which we can find because the $S^3$ is a calibrated cycle. Recalling
that $\Phi\sim \Phi_0 + tdf \wedge \omega$, we get:
\be {\rm vol} \ = \  \int_{S^3} \Phi \ \sim\  \int t\del_\lambda f
\,d\lambda \sim t\,(f(q) -f(p)) \ee
This agrees with the action of the gradient flow instanton. Clearly
this generalizes to more complicated ALE fibrations. Note that we
should not interpret these $M2$ instantons as saying that our gauge
theory breaks down, we are still describing a tunnelling effect in
the $7d$ gauge theory.

We can also interpret these instantons in the spectral cover
picture. It will not come as a surprise that they lift to disk
instantons. To see this, consider a disk stretching between the zero
section and the Lagrangian brane defined by the function $f$, which
projects to the gradient flow trajectory on $S^3$ between the
critical points $p$ and $q$. Its area, measured with respect to the
standard symplectic form on $T^*S^3$, is given by
\be \int_{-\infty}^{\infty} d\lambda \int_0^1 dx \, x |t\nabla f| =
\half t(f(q) - f(p)) \ee
and therefore the action of the disk instanton also agrees with the
action of the gradient flow instanton.

 Now the beauty of the Morse
complex is that it reconstructs the ordinary cohomology of the
underlying manifold, i.e. we have
\be H^*_{Morse}(M,f) = H^*(M, {\bf R}) \ee
One way to understand this isomorphism is by noticing
\cite{Witten:1982im} that our differential operator is related to
the ordinary exterior differential by a similarity transformation
\be d_t = e^{-tf}d e^{tf} = d + t\,df \wedge \ee
Thus the betti numbers are independent of $t$. In the limit $t\to
\infty$ we obtain the Morse complex, and in the limit $t\to 0$ we
recover the definition of the ordinary de Rham cohomology.

There is also a version of this isomorphism for manifolds with
boundary, which is the case relevant for us. Recall that we
effectively have boundaries where the Morse function (or
electro-static potential) becomes infinite. We will assume the
boundary may be split up into disjoint positively charged and
negatively charged pieces
\be \del M = \del^+ M \cup \del^-M \ee
Then the Morse complex reconstructs the relative cohomology
\be H^*_{Morse}(M,f) = H^*(M,\del^+M). \ee
Actually in order for Morse theory with boundaries to be
well-defined and reproduce the relative cohomology, we must ensure
that the gradient flow trajectories connecting critical points do
not hit the boundary. This is automatically the case for the
harmonic solutions that satisfy the $D$-term equations. In the
electro-statics analogy, the critical points are the points where a
test charge would experience zero force. The gradient flow
trajectories describe the possible trajectories of a positive test
charge. A trajectory connecting two critical points cannot hit a
boundary -- the potential energy is minimized at a boundary and the
test charge cannot climb back out.

Therefore we conclude that the massless matter content depends only
on topological properties of the configuration and is independent of
the explicit solution to the $D$-term equation. The massless matter
content is simply given by the ranks of the relative cohomology
groups:
\be N_\chi(R) = h^1(M, \del^+M), \qquad N_{\bar{\chi}}(R) =
h^2(M,\del^+M) \ee
where $\del^+M$ is the boundary of $M$ where the Morse function is
increasing, i.e. where the positive charges are located. Therefore
the computation of the spectrum is reduced to a purely {\it
combinatorial} problem involving triangulations or cell complexes.
In particular the net number in the representation $R$ is simply
given by
\be {\rm net \ chiral}(R) \ =\ \chi(M,\del^+M) \ee
As we will discuss later, this formula holds much more
generally.

More generally we may have multiple abelian Higgs fields. Suppose
there are two, with associated Morse functions $f_a,f_b$. Let us
denote the boundaries as
\be \del_a = \del^+_a - \del^-_a \ee
where we used the minus sign to indicate that there are negative
charges located on $\del^-_a$. We will assume again that positive
and negative boundaries do not intersect, and we will also assume
for convenience that the charge density along the boundary is
uniform. Then it's not hard to see that for a general linear
combination $q_a f_a + q_b f_b$, the boundary is given by
\be \del_{q_a,q_b} = q_a \del^+_a -q_a \del^-_a + q_b \del^+_b - q_b
\del^-_b \ee
Depending on how the Higgs fields are embedded into a non-abelian
group $G$, one will be interested in the critical points of various
linear combinations of $f_a$ and $f_b$. These linear combinations
depend on the $U(1) \times U(1)$ charges $(q_a,q_b)$ of the
multiplets appearing in the decomposition of the adjoint of $G$
under $H \times U(1)_a \times U(1)_b$. In each case we have
\be N_\chi(q_a,q_b) = H^1(M,q_a f_a + q_b f_b) =
H^1(M,\del^+_{q_a,q_b}) \ee

So far, our results were stated as certain properties of the Higgs
field. We may also restate some of the results in the spectral cover
description, which yields a more geometric picture. We have already
seen that critical points correspond to intersections of components
of the spectral cover with the zero section. We have also seem that
gradient flow trajectories lift to disk instantons in $T^*S^3$.
Therefore the instanton corrected spectrum is computed by the Floer
cohomology groups
\be HF^*(C_0,C_E) \ee
where we used $C_0$ to denote the zero section. Indeed it is well
known in the literature that Floer cohomology in the cotangent
bundle coincides with Morse/Novikov cohomology on the base manifold
\cite{FloerLag,Witten:1992fb,FukayaMorse}.

The analysis of the charged chiral spectrum implies that if $\#
\,{\rm Crit}_i(f)>h^i(M,\del^+ M)$, then there are chiral fields in
the spectrum whose masses are exponentially suppressed compared to
the GUT or KK scale:
\be M^2\ \sim\  \ e^{-{1\over { \alpha_{GUT}}}}M_{GUT}^2 \ee
These massive modes modify the running of the gauge couplings below
the GUT scale (which may be identified with the KK scale up to
threshold corrections), and may provide channels for proton decay.
The GUT breaking mechanism of \cite{Witten:2001bf} using discrete
Wilson lines in fact {\it requires} the existence of such relatively
light massive modes, namely the Higgsino triplets, and depending on
the model there could be additional modes. The GUT breaking
mechanism of \cite{Witten:2001bf} in principle also allow one to
eliminate dimension five operators leading to proton decay. However
with such a low mass for the triplets, even dimension six proton
decay could lead to trouble. Therefore it was suggested in
\cite{Friedmann:2002ty} that for the GUT breaking mechanism in
\cite{Witten:2001bf} to be viable, one should also modify the
running so as to get $\alpha_{GUT} \sim 0.2-0.3$. This could be
engineered by having additional $\bfv\,\bfb$ pairs below the GUT
scale. We will briefly suggest a different GUT-breaking mechanism in
section \ref{GUTbreaking}.

\newsubsection{Comments on anomaly cancellation}

In type IIa string theory, non-abelian anomalies due to chiral
fermions at brane intersections can be cancelled by anomaly inflow.
To see this, we have an interaction
\be\label{D6CScoupling} S_{CS}= \int C^{(1)} \wedge {\rm Tr}(F^3) =
-\int F^{(2)} \wedge \omega_5(A) \ee
Under a gauge transformation $\Lambda$ we have
\be \delta_\Lambda S_{CS} = -\int F^{(2)}\wedge \delta_\Lambda
\omega_5(A) = \int dF^{(2)} \wedge I_4^1(A,\Lambda) \ee
Further, $6$-branes are monopole configurations for $C^{(1)}$:
\be dF^{(2)}/2\pi \ \sim\ \sum_a N_a \delta^3(P_a) \ee
Therefore the term $\delta_\Lambda S_{CS}$ above is of precisely the
right form to cancel non-abelian anomalies due to chiral fermions at
brane intersections.

In $M$-theory, the RR gauge field $C^{(1)}$ is generally massive,
and so is not included as a propagating degree of freedom in the
effective action. However after integrating it out, there must be a
residual interaction which is not invariant under gauge
transformations, for otherwise the anomalies above could not be
cancelled. Reference \cite{Witten:2001uq} explained what this
residual interaction looks like in a local model. The geometry near
a 6-brane locally looks like an $A_n$ ALE fibration over $S^3$.
The ALE has a natural $U(1)$ isometry that commutes with the
holonomy, and as we fiber over $S^3$ we may get a non-trivial $U(1)$
bundle. The curvature of the corresponding $U(1)$ bundle over $S^3$
is denoted by $K$; it becomes $F^{(2)}$ in a IIa limit. Then in
\cite{Witten:2001uq} it is essentially argued that the coupling
(\ref{D6CScoupling}) survives in $M$-theory in the following form:
\be \int d^7 x {K\over 2\pi}\wedge \omega_5(A), \qquad dK = \sum n_a
\delta_{P_a} \ee
where $P_a$ are the locations of chiral and anti-chiral matter, and
$n_a$ their multiplicities. As written, this interaction only makes
sense for local geometries which are fibered by $A_n$ ALE spaces and
not for more general $G_2$-manifolds, but presumably there is a more
general expression that reduces to the above one in a scaling limit.
This is an additional coupling in the action beyond the terms we
have considered so far, which is needed for consistency. The
associated tadpole constraint expresses cancellation of the
non-abelian anomalies.

One may also consider abelian anomalies. Let us briefly review some
arguments in \cite{Witten:2001uq} (see also \cite{Bilal:2003pz}).
Chiral fields are localized at codimension 7 singularities, and
locally the $G_2$ metric may be written as
\be
ds^2 \sim dr^2 + r^2 ds_{Y_\alpha}^2 \ee
where $r$ is a radial coordinate. We cut such a local neighbourhood
around each codimension 7 singularity. Then our $G_2$ manifold $X_7$
becomes a manifold-with-boundary $X_7'$, with $\del X_7' =
\cup_\alpha Y_\alpha$. We expand $C_3$ in harmonic forms on $X_7'$:
\be C_3\ \sim\ A_i \wedge \omega^i + a_j \wedge \chi^j\ee
Here $\omega^i$ are harmonic 2-forms and $\chi^j$ are harmonic
3-forms, so the $A_i$ are our four-dimensional abelian gauge fields
and the $a_j$ are four-dimensional axions. Then under a gauge
transformation $C_3 \to C_3 + d \Lambda$, the Chern-Simons term
varies as
\be \delta_\Lambda S_{CS} \sim -\int_{R^4 \times \del X} \Lambda
\wedge G \wedge G \ \sim\ -\sum_\alpha \int_{R^4} \epsilon_i\,  F_j
\wedge F_k \ \int_{Y_\alpha} \omega^i \wedge \omega^j \wedge
\omega^k \ee
where we decomposed $\Lambda$ as $\Lambda \sim \epsilon_i\,
\omega^i$. This is cancelled if there are fermion zero modes
$\psi_\sigma$ localized at the singularity, with $U(1)$ anomalies
given by
\be \sum_\sigma q^i_\sigma q^j_\sigma q^k_\sigma\ =\ \int_{Y_\alpha}
\omega^i \wedge \omega^j \wedge \omega^k \ee
Moreover from Stokes's theorem we get
\be 0\ =\ \sum_{\sigma, \alpha} q^i_\sigma q^j_\sigma q^k_\sigma \ee
which one interprets as cancellation of the cubic $U(1)$ anomalies.
Similarly one may discuss mixed abelian-gravitational anomalies and
mixed abelian/non-abelian anomalies.

At first sight there is one puzzling aspect about this derivation.
It implies that in $M$-theory compactifications on $G_2$ manifolds,
the chiral spectrum is such that the light $U(1)$'s are always
non-anomalous, and no Green-Schwarz mechanism is ever needed. On the
other hand, type IIa with $D6$ branes lifts to $M$-theory on $G_2$,
and there can also be heterotic duals. In both of these contexts,
there can be light anomalous $U(1)$'s and there is a Green-Schwarz
mechanism for their cancellation.

The likely resolution to this puzzle is as follows. In these other
settings, the anomalous $U(1)$'s obtain a mass of order $\sim
g_{YM,4}\ell_s^{-1}$ through the Green-Schwarz terms, and this mass
is always parametrically lighter than the KK scale. Thus it makes
sense to have anomalous $U(1)$'s below the KK scale and a
Green-Schwarz mechanism for cancelling their anomalies. This is true
even in $F$-theory. However the lift from IIa to $M$-theory is a
little more subtle, and it seems a priori possible that such
anomalous $U(1)$'s in type IIa will lift to massive $U(1)$'s in
$M$-theory with masses scaling like $1/R_{KK}$. In that case we
should treat these massive $U(1)$'s on the same footing as other
massive $U(1)$ gauge bosons, and there will be no anomalous $U(1)$'s
in the effective action below the KK scale, in agreement with the
arguments of \cite{Witten:2001uq}.

\newsubsection{Techniques from algebraic topology}
\subseclabel{AlgTop}

\begin{figure}[t]
\begin{center}
            \scalebox{.9}{
               \includegraphics[width=\textwidth]{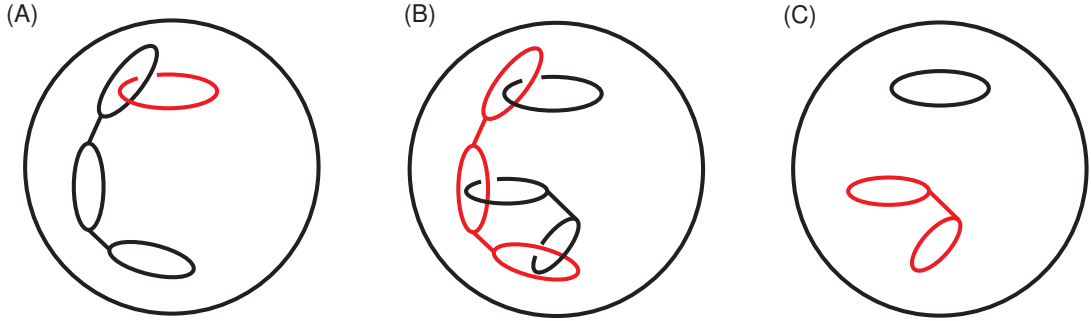}
               }
\end{center}
\vspace{-.5cm}
\begin{center}
\parbox{14cm}{\caption{ \it Positively charged (red) and negatively charged
(black) boundaries of Morse functions on $S^3$. Case (a) yields an
electro-static potential that we call $f_6$, case (c) yields the
potential $f_1$, and case (b) corresponds to the linear combination
$-f_1 - f_6$. Note that we took one of the pieces of the boundary common
to $f_1$ and $f_6$, but with the charge density for $f_6$ dominant
there.}\label{ToyGut}}
\end{center}
 \end{figure}

 In this section we would like to apply some
simple techniques from algebraic topology in order to compute the
relative betti numbers $h^i(M,\del^+ M)$ in terms of the topological
properties of the positive and negative boundaries. We actually work
with the homology, which can easily be dualized to cohomology. We
assume that the positive charge density is smeared along a graph
$\Delta^+$ with $n_+$ components and $\ell_+$ loops, and similarly
the negative charge density is smeared along a graph $\Delta^-$ with
$n_-$ components and $\ell_-$ loops. For some examples see figure
\ref{ToyGut} in section \ref{Example}. The open manifold $M$ is
identified with $S^3\backslash (\Delta^+\cup \Delta^-)$.

 In order to compute the relative betti numbers and the
Euler character, there are two relevant long exact sequences. The
first is the Mayer-Vietoris sequence. Suppose a manifold $X$ is
covered by two open sets $U,V$. Then we have the long exact sequence
\be \ldots \to H_i(U\cap V) \to H_i(U) \oplus H_i(V) \to H_i(X) \to
H_{i-1}(U \cap V) \to \ldots \ee
In particular, the Euler characters are related as
\be \chi(X) = \chi(U) + \chi(V) - \chi(U\cap V) \ee
In the application we have in mind, $X=S^3$, $U$ is $S^3$ with a
small tubular neighbourhood of the negative boundary excised, and
$V$ is itself a tubular neighbourhood of the negative boundary.
Suppose the graph has $n_-$ components and has $\ell_-$ loops. Then
$U\cap V$ is topologically a collection of higher genus Riemann
surfaces, so we have $b_i(U\cap V) = \{n_-,2\ell_-,n_-,0\}$.
Furthermore we have $b_i(V) = b_i(\Delta^-)$. It follows from the
long exact sequence that
\be b_3(S^3\backslash \Delta^-) =0, \quad b_2(S^3\backslash
\Delta^-) = n_- -1, \quad b_1(S^3\backslash \Delta^-) = \ell_-,
\quad  b_0(S^3\backslash \Delta^-) = 1.\ee
In particular
\be \chi(S^3\backslash \Delta^-) = n_- -\ell_- \ee

The second sequence we need is the one for relative homology. If $A$
is a subset of $X$, then we have
\be \ldots \to H_i(A) \to H_i(X) \to H_i(X,A) \to H_{i-1}(A) \to
\ldots \ee
In particular
\be \chi(X,A) = \chi(X) - \chi(A) \ee
For our application, we would like to take $X$ to be $M \sim
S^3\backslash (\Delta^+\cup\Delta^-)$, and $A$ to be a small tubular
neighbourhood of $\Delta^+\sim \del^+M$. In fact by the excision
axiom, because $\Delta^+ \subset A$, it is equivalent to take $X =
S^3\backslash \Delta^-$. The precise betti numbers $b_1(X,A)$ and
$b_2(X,A)$ depend on the details of how the graphs are linked. From
the long exact sequence, we find that they are given by
\be
\begin{array}{rl}
  b_2(M,\del^+M) =  &   n_--1+\ell_+ - r \\ [2mm]
  b_1(M,\del^+M) = & n_+ -1 + \ell_-  - r
\end{array}
  \qquad \qquad 0\leq r \leq {\rm min}(\ell_+,\ell_-) \ee
Here $r$ is  the rank of the inclusion map
\be H_1(\del^+M) \to H_1(S^3\backslash \Delta^-) \ee
That is, any loop in $\Delta^+$ is naturally embedded in
$S^3\backslash \Delta^-$, and $r$ counts the number of loops that
remain independent in homology after embedding. Thus we see that $r$
indeed describes the linking between the positive and negative
graphs. If the graphs are unlinked then $r=0$. Moreover the Euler
character is easily calculated and given by
\be \chi(M,\del^+M) = \chi(M) - \chi(\del^+M) = n_- -\ell_- -n_+ +
\ell_+  \ee
which is independent of $r$. Thus the net number of generations is
easily computed from the topology of the positive and negative
boundaries.

\newsubsection{Example}
\subseclabel{Example}

 \begin{figure}[t]
\begin{center}
            \scalebox{.4}{
               \includegraphics[width=\textwidth]{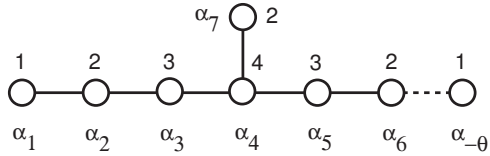}
               }
\end{center}
\vspace{-.5cm}
\begin{center}
\parbox{14cm}{\caption{ \it The extended $E_7$ Dynkin diagram and Dynkin indices.}\label{E7Dynkin}}
\end{center}
 \end{figure}

In this section we would like to consider a toy GUT model, obtained
by compactifying a gauge theory with gauge group $E_7$ on the
three-manifold $Q_3 = S^3$. We will use the labelling of the roots
shown in figure \ref{E7Dynkin}. The group $E_7$ contains a maximal
$SO(10) \times U(1)_a \times U(1)_b$ subgroup, and we can take the
$SO(10)$ to be generated by $\{ \alpha_2,\alpha_3,\alpha_4,
\alpha_5,\alpha_7 \}$. Under this subgroup, the adjoint
representation of $E_7$ decomposes as
\ba {\bf 133} &=& {\bf 45}_{0,0} + {\bf 1}_{0,0}+ {\bf 1}_{0,0} +
{\bf r} +  \bar{\bf r} \eol {\bf r} & =& {\bf 16}_{-2,1} + {\bf
16}_{0,-3} + {\bf 10}_{2,2} + {\bf 1}_{-2,4} \ea
In terms of our canonical basis $\omega_i$ dual to $\alpha_j$, the
two $U(1)$'s correspond to
\be \omega_a = 2\omega_1, \qquad \omega_b = 2\omega_1 - 3 \omega_6
\ee

We would like to have an unbroken $SO(10)$ gauge group after
compactification. In order to achieve this breaking pattern, we need
to turn on a profile for the Higgs fields corresponding to $U(1)_a$
and $ U(1)_b$. The Morse functions will be denoted as $\log h_1
=f_1$ and $\log h_6 = f_6$. That is, the FI parameters are given by
\be \lambda_1 = df_1, \qquad \lambda_6 = df_6 , \qquad
\lambda_\theta = -df_1 - 2 df_6 \ee
Then the various chiral fields are counted by the following Morse
cohomology groups:
\be \begin{array}{rclrcl} {\bf 16}_{0,-3} &\to & h^1(M,f_6) \qquad &
{\bf 16}_{-2,1} &\to & h^1(M,-f_1-f_6) \eol [1mm]  \bt_{2,2}\ &\to &
h^1(M,f_1) \qquad & {\bf 1}_{-2,4} &\to & h^1(M,-f_1-2f_6)
\end{array}
\ee
Similarly the chiral fields in the conjugate representations are
counted by $h^2(M,f)$.

Now we will define the electro-static potentials $f_1$ and $f_6$ by
specifying suitable positive and negative charges on $S^3$. An
example is shown in figure \ref{ToyGut}. To compute the spectrum, we
use the formulae derived in section \ref{AlgTop}, dualized to
cohomology:
\be
\begin{array}{rl}
  h^1(M,\del^+M) =  &  n_+ -1 + \ell_-  - r  \\ [2mm]
  h^2(M,\del^+M) = & n_--1+\ell_+ - r
\end{array}
  \qquad \qquad 0\leq r \leq {\rm min}(\ell_+,\ell_-) \ee
Using the charge configurations in figure \ref{ToyGut}, we find
\be
\begin{array}{rclrclrcl}
h^1(M,f_6) &=& 2 \qquad& h^1(M,-f_6-f_1) &=& 1 \qquad & h^1(M,f_1) = 1  \eol
h^2(M,f_6) &=& 0 & h^2(M,-f_6-f_1) &=& 0 & h^2(M,f_1) = 2
\end{array}
\ee
Therefore we have precisely three chiral generations in the ${\bf 16}$,
three Higgs fields, and also one chiral field in the representation
${\bf 1}_{-2,4}$. The Yukawa coupling
\be {\bf 16}_{0,-3} \times {\bf 16}_{-2,1} \times {\bf 10}_{2,2} \ee
is in principle allowed by the symmetries.

This example is unrealistic in a number of ways. In our non-compact
set-up the $U(1)_a\times U(1)_b$ is not dynamical, but it imposes
some strong selection rules. Since the $U(1)$'s are anomalous under
the spectrum of the local model, we expect them to become massive
upon compactification, but interactions violating the selection
rules would then still be supressed by the compactification scale. A
nicer way would be to break the extra $U(1)$'s by using non-abelian
Higgs fields. Or one could start with an abelian configuration and
try to turn on an expectation value for the ${\bf 1}_{-2,4}$. In the
present example such a VEV would still leave an unbroken $U(1)$, so
it would be better to start with $E_8$ instead since this can be
broken to $SO(10)$ by turning on an $SU(4)$ valued Higgs field.
Finally we haven't yet incorporated a mechanism for breaking the GUT
group to the Standard Model.

\newsubsection{Breaking the GUT group with an abelian Higgs field}
\subseclabel{GUTbreaking}

In order to break the GUT group, we can in principle proceed in two
ways. One method is to engineer a four-dimensional Higgs field that
can do the breaking, like an adjoint field. This is not possible if
we work with $Q_3 = S^3/\Gamma$ since $h_1(S^3/\Gamma) = 0$, and
even if it were one would end up with a conventional
four-dimensional GUT model with the associated problems like
doublet-triplet splitting.

The other method is to give a VEV to a charged field in the higher
dimensional gauge theory in the process of compactification. The
available charged fields which can get a VEV are the $7d$ gauge
field and the $7d$ Higgs field.

We can turn on a VEV for the gauge field in two ways. If
$\pi_1(Q_3)$ is non-zero we can use discrete Wilson lines with a
hypercharge component. This is the conventional mechanism used in
the heterotic string and introduced in the $M$-theory context in
\cite{Witten:2001bf}. Unfortunately as we already discussed, this
mechanism leads to light Higgsino triplets and looks somewhat less
than desirable. The other possibility is to turn on a flux with a
hypercharge component. However we have $h^2(S^3/\Gamma)=0$ so this
is not available for us \cite{Donagi:2008kj}.

There is then still one other option: we could turn on an abelian
hypercharged Higgs field to break $SU(5) \to SU(3) \times SU(2)
\times U(1)$. This corresponds to a reducible spectral cover. Apart
from the usual Standard Model fields which descend from the $\bt$
and $\bfb$ however, there can be additional unwanted matter. Let us
denote the pair $(C,L)$ simply by ${\cal L}$. The spectral cover
moduli decompose as
\be HF^1({\cal L},{\cal L}) = \sum_{m,n} HF^1({\cal L}_m,{\cal L}_n)
\ee
where $m,n$ run over the irreducible components. The off-diagonal
components are charged under the extra $U(1)$'s, which includes
hypercharge in this case. We saw an example of this in section
\ref{Example}, where we had a scalar ${\bf 1}_{-2,4}$. So in this
scenario one has to make sure that such hypercharged moduli are
massive, i.e. there is no deformation to a smooth spectral cover,
because there are no light hypercharged scalars in the real world.
This can likely be arranged by a suitable boundary condition or by
turning on flat spectral line bundles on the irreducible components
of the cover which can not be obtained as a limit of a smooth line
bundle after deformation. The crucial part is to check that this can
be done while still satisfying the $D$-terms.

Another possible issue is that in the local set-ups we have been
discussing, the $U(1)$ would not be dynamical. It can become
dynamical when embedded in a compact model but it may have some UV
sensitivity. Nevertheless this would lead to very different
signatures than breaking by discrete Wilson lines, and so may be
worth pursuing.

\newsubsection{Superpotential terms}

In this section we would like to explain how superpotential
couplings are recovered from Morse theory. In the $1/t$ expansion we
can think of such couplings as generated by membrane instantons or
disk instantons, which map to trees of gradient flow trajectories on
$S^3$. The prescriptions were originally found by
\cite{FukayaMorse,BetzCohen}. We focus on the Yukawa couplings and
quartic couplings. This is all that is needed for practical
purposes. The Yukawa couplings correspond to the cup product on
cohomology, and the quartic couplings to the Massey product, and so
once more we see that the $F$-term data reflects the underlying
topology and may be computed using alternative, combinatorial
methods.

To define a three-point function we need three functions $f_i, i =
1,2,3$ such that the differences $f_{ij} = f_i - f_j$ are Morse
functions. In practice one of the $f_i$ will correspond to the zero
section and so is taken constant. Our chiral fields correspond to
certain linear combinations of index one critical points of the
$f_{ij}$. We assume that a chiral field can be associated to a
definite critical point. One can extend the definition of the
three-point function using linearity.

 \begin{figure}[b]
\begin{center}
            \scalebox{.2}{
               \includegraphics[width=\textwidth]{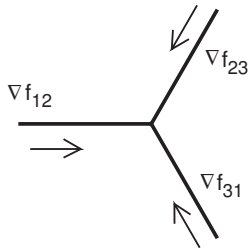}
               }
\end{center}
\vspace{-.5cm}
\begin{center}
\parbox{14cm}{\caption{ \it Graph of gradient flow trajectories relevant for
Yukawa couplings.}\label{MorseYukawa}}
\end{center}
 \end{figure}

The three-point function is given by the classical overlap of the
wave functions of the chiral zero modes. In the $1/t$ approximation,
it receives contributions from minimal area membrane instantons,
which map to graphs of gradient flow trajectories in $S^3$ along the
lines we discussed in section \ref{MorseCohomology}. Namely we
consider embedding graph as in figure \ref{MorseYukawa} into $S^3$,
such that the edges get mapped to gradient flows of $f_{ij}$ as
indicated, and the ends get mapped to an index one critical point of
$f_{ij}$. The moduli space of such graphs is denoted by
$M(p_{12},p_{23},p_{31})$. When non-empty, it is a manifold of
dimension
\be \dim M(p_{12},p_{23},p_{31}) = m(p_{12}) + m(p_{23}) +
m(p_{31})-3 \ee
In particular when $m(p_{12}) = m(p_{23}) = m(p_{31}) = 1$, it is a
finite set of points counted with signs. The action of such an
instanton is given by
\be \exp -|f_{12}(p_{12}) +f_{23}(p_{23}) +f_{31}(p_{31})| \ee
and so the Yukawa coupling is given by
\be \lambda_{123} = \sum_{graphs} n(p_{12},p_{23},p_{31})\, e^{
-|f_{12}(p_{12}) +f_{23}(p_{23}) +f_{31}(p_{31})|} \ee
Although the exponential factors may be scaled out by field
redefinitions and are therefore ignored in the mathematics
literature, they are physically relevant because the field
redefinitions would change the normalization of the kinetic terms.

 \begin{figure}[t]
\begin{center}
            \scalebox{.8}{
               \includegraphics[width=\textwidth]{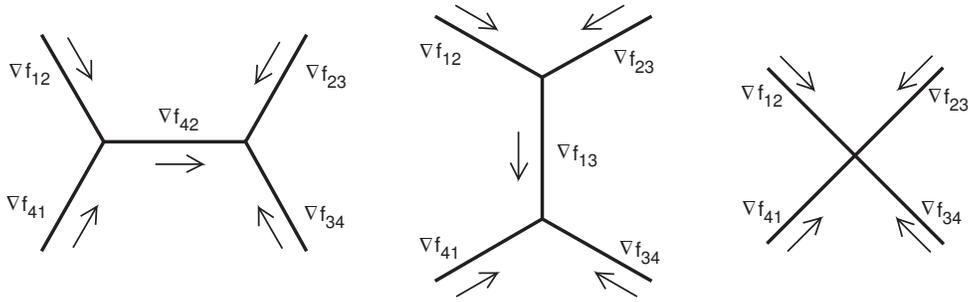}
               }
\end{center}
\vspace{-.5cm}
\begin{center}
\parbox{14cm}{\caption{ \it Graphs of gradient flow trajectories
generating quartic couplings.}\label{MorseQuarticW}}
\end{center}
 \end{figure}

Similarly we may recover the four-point function. In this case use
four functions $f_i, i = 1,2,3,4$ and we look for graphs of gradient
trajectories as in figure \ref{MorseQuarticW}. The virtual dimension
of the moduli space of graphs of type (a) and (b) is given by
\be \dim M(p_{12},p_{23},p_{34},p_{41}) = m(p_{12}) + m(p_{23}) +
m(p_{34}) + m(p_{41}) - 4 \ee
These two moduli spaces are glued along the moduli space of graphs
of type (c), whose dimension is one lower. Again we are interested
in the case when all $m(i) = 1$, in which case we can count a
discrete number of graphs of type (a) and (b). The quartic coupling
is defined as
\be \lambda_{1234} = \sum_{graphs} \# M(p_{12},p_{23},p_{34},p_{41})
\,e^{-|\sum_i f_{i,i+1}(p_{i,i+1})|} \ee
There exist alternative ways to compute the four-point function.
Once the boundary operator and the three-point functions have been
obtained, one may also construct higher point functions through a
recursion relation \cite{Witten:1992fb,Merkulov}. This may be
thought of as a Feynman diagram expansion of our Chern-Simons theory
(\ref{complexCS}), however it is more general in that it does not
necessarily require that we use the Morse-Smale complex. In case of
the four-point function this computes the (lenght 3) Massey product.
If the boundary operator and the three-point function have been
normalized correctly, these two constructions should be equivalent
up to field redefinition of the form $\Phi_i \to \Phi_i +
c_{ijk}\Phi_j\Phi_k + $higher order. There are generalizations to
higher point functions along the same lines.

\newsubsection{Unsuppressed couplings and degenerate critical points}

So far we have assumed that chiral fields are localized at
non-degenerate critical points. In this case the interactions
between chiral fields are described as being generated by instanton
effects. Interactions are therefore generically small, and this will
lead to phenomenological problems. In particular, this would lead to
the prediction that the top quark Yukawa coupling is rather small,
whereas in fact it turns out to be of order one.

In order to get unsuppressed couplings, we must tune the boundary
conditions on the flavour branes so that the endpoints of the
trivalent graph for the Yukawa coupling live close together. In this
limit, the instanton approximation is not good, but we can get
another weakly coupled description by analyzing the Dirac equation
and couplings directly for a degenerate (non-Morse) critical point.
Although we have not carried out the analysis, one expects to be
able to get a non-zero classical contribution to the Yukawa
couplings this way, at least for the down type or off-diagonal
up-type Yukawa couplings.

This however may still not be enough for the top quark, as was
pointed out in \cite{Tatar:2006dc}. In our present language, if the
$Q_L$ and $U_R$ come from the same critical point, then the
trivalent graph for the Yukawa coupling must have two legs on that
critical point. This seems impossible to achieve if the graph must
live in a small local neighbourhood of the critical point, because
the two $\bt$'s in the $\bt \cdot \bt\cdot \bfv$ coupling have
different weights under the holonomy group of the Higgs bundle (i.e.
they live on different sheets of the spectral cover), and so one of
the legs of the graph will have to pass through a branch point or
wrap around a boundary. Thus no matter how close the $\bt$ and
$\bfv_h$, the diagonal up-type Yukawa couplings would still seem to
be subject to some exponential suppression. Overcoming this issue is
one of the key points that needs to be addressed in $M$-theory
phenomenology.

It's also interesting to point out that one needs exceptional gauge
groups in $7d$ to describe the top quark Yukawa coupling in
$M$-theory, just as in $F$-theory or the heterotic string, and
essentially for the same reason \cite{Tatar:2006dc}. In the language
of the present paper, this is because the fermion zero modes live in
non-trivial representations of the holonomy group of the Higgs
bundle, which is itself a subgroup of the (complexified) gauge group
in $7d$. Interaction terms must be invariant however, so it must be
possible to make the Yukawa coupling a singlet under this holonomy
group. In the case of the top quark Yukawa coupling in an $SU(5)$
GUT model, this singles out the exceptional groups for purely group
theoretical reasons.

In $F$-theory one may take a scaling limit to map models to a IIb
description \cite{Donagi:2009ra}. The top quark Yukawa is then
described as a $D1$ instanton effect. An analogous limit is not yet
known for $G_2$ manifolds, but we know that membrane instantons get
mapped to worldsheet disk instantons and $D2$-instantons. As usual,
the top quark Yukawa is charged under the $U(1) \subset U(5)$ in the
IIa description. This $U(1)$ is anomalous but it is parametrically
lighter than the KK scale in the IIa limit. The instanton action
will then have to shift under the $U(1)$ gauge transformation in
order to compensate for the lack of invariance of the up type Yukawa
coupling, or it should be neutral for the case of the down type
coupling. Since the Ramond three-form shifts under such a gauge
transformation and the $NS$ two-form doesn't, we expect that
membrane instantons get mapped to $D2$ instantons if they generate
an up type Yukawa, and to disk instantons if they generate a down
type Yukawa.

\newsubsection{Novikov cohomology}

Now we discuss the more general possibilities for the boundary
behaviour of abelian Higgs fields:
\ba A&\sim &\alpha d\theta \eol \phi &\sim& \beta d\log r + \gamma
d\theta \ea
First we consider $\beta,\gamma \not = 0$ but still $\alpha=0$. In
this case we can no longer define an electro-static potential, since
it would have to be multi-valued. Related to this, there may be an
infinite number of gradient flow trajectories connecting two
critical points, and hence the boundary operator is no longer
well-defined. However we can still define a potential on a multiple
cover of $M$, define critical points and flow lines, and define a
equivariant boundary operator and a cohomology theory. This more
general cohomology is called Novikov cohomology.  We will next
explain how this arises from the supersymmetric quantum mechanics
discussed previously, which can actually be defined by replacing
$df$ by an arbitrary closed one-form $\phi$.

The Novikov homology depends on the cohomology class $[\phi]$ of the
closed one-form, which specifies the monodromies. The $D$-terms
imply that $\phi$ is harmonic. By Hodge theory arguments, there
should be a unique harmonic form in the class whose periods are
specified by $[\phi]$. Given a cohomology class, we may construct
the minimal cover $\tilde{M}_\phi$ on which $\phi$ becomes exact,
$\pi^*\phi = df$. Generically this is the universal cover. Denote by
$\Gamma$ the group of covering transformations. Then $f$ satisfies
$f(gx) = f(x) + \xi(g),\ g \in \Gamma$, where $\xi(g)$ are the
`periods' of $[\phi]$. Note that $\xi(g)$ is independent of $x$.

Since locally the situation is exactly the same as when the periods
of $\phi$ vanish, we still get a chiral fermion for each critical
point of $\phi$ in the $1/t$ expansion. Thus the generators of the
Novikov cohomology are still given by the critical points. However
the coefficients will no longer be real-valued but rather valued in
a power series in order to keep track of additional information. As
before the action of $d_t$ defines a boundary operator
\be d_t \ket{a} = \sum_{b\in {\rm Cr}_{i+1}(M), g\in \Gamma}
n(a,gb)e^{-t(f(gb) - f(a))} \,  \ket{b} \ee
Note that $n(a,gb)$ is only non-zero when $f(gb)>f(a)$. By rescaling
$\ket{a},\ket{b}$ we can write this as
\be d_t \ket{a} = \sum_{\ket{b}\in {\rm Cr}_{i+1}(M), g\in \Gamma}
n(a,gb)q^{\xi(g)} \, \ket{b} \ee
where $q = e^{-t}$ and $q^{\xi(g)}$ corresponds to the part of the
instanton action that cannot be scaled out.

We see that the `numbers' multiplying $\ket{b}$ are no longer
integers, but live in a ring of power series called the Novikov ring
$Nov(\Gamma)$. It is defined as the ring of formal power series
\be Nov(\Gamma) = \left\{\sum_{i=0}^\infty n_{i} q^{\gamma_i}\ |\
\gamma_i \in {\bf R}, \gamma_i < \gamma_{i+1}, \gamma_i \to
-\infty\right\} \ee
i.e. $\gamma_i$ takes values in a discrete set which tends to minus
infinity, and the $n_i$ are integers. The cohomology of $d_t$ is
called the Novikov cohomology:
\be H^*_{Nov}(M,[\phi])\ 
\ee
Since the chain complex is a module over $Nov(\Gamma)$, hence so is
the cohomology. It is known that $H^*_{N}(M,[\phi])$ decomposes as a
finite sum of free and torsion modules over $Nov(\Gamma)$.

The zero modes we are after are annihilated by both $d_t$ and
$d_t^\dagger$. This number may jump for a finite number of values of
$t$, but excluding those values it is independent of $t$. Hence the
generic number of zero modes corresponds to the number of generators
of the cohomology of $d_t$ over the Novikov ring, i.e. by the
Novikov-Betti numbers:
\be h^i(M,[\phi]) = {\rm rank}_{Nov(\Gamma)}\ H^i_N(M,[\phi]) \ee
If $[\phi] = 0$, i.e. if $\phi = df$ for some $f$, this reduces to
the usual Betti numbers. Since our Dirac equation is defined over
the real numbers, we can ignore the torsion. Thus $h^i(M,[\phi])$
counts the massless chiral matter.

It is natural to expect that the effect of turning on additional
monodromy is to reduce the total amount of massless matter, i.e. we
expect
\be h^i(M,[\phi]) \leq h^i(M,[\phi] = 0) \ee
In the case without boundary, a proof can be found in \cite{Farber}.
Furthermore, if the periods can be turned off continuously (as would
certainly be the case if we work on a simply connected manifold like
$S^3$), then the net amount of chiral matter can clearly not change.
Thus
\be h^2(M,[\phi]) -h^1(M,[\phi]) \ =\ \chi(M,\del^+ M) \ee
As we vary $[\phi]$ in $ H^1(M,{\bf R})$, the Betti numbers
$h^i(M,[\phi])$ are generically constant, but may jump up on
algebraic subsets of $H^1(M,{\bf R})$. However they must always
satisfy the identities above.

Finally we would like to also allow for $\alpha \not = 0$, i.e. we
turn on a flat spectral line bundle. In this case we also need to
take into account the holonomy of the gauge field, and the action of
$d_t$ is modified to
\be d_t \ket{a} = \sum_{b\in {\rm Cr}_{i+1}(M), g\in \Gamma}
n(a,gb)\, e^{i\int_{a}^{gb}\! A}\, e^{-\half t(f(gb) - f(a))} \,
\ket{b} \ee
Again after scaling out a common piece, we can write this as
\be d_t \ket{a} = \sum_{b\in {\rm Cr}_{i+1}(M), g\in \Gamma}
n(a,gb)\,  e^{ i\rho(g)} q^{\xi(g)} \, \ket{b} \ee
Here $\rho(g) = \int_b^{gb}A$ is the representation describing the
holonomies of the flat line bundle. This yields a twisted version of
Novikov cohomology. Not much seems to be known about it. However
when $\gamma \to 0$ it reduces to the cohomology with values in $L$,
i.e. $H^i(M,L)$ \cite{Farber}. Since we would expect that turning on
$\gamma$ generally only decreases the betti numbers, this can be
used to get even stronger constraints on the spectrum. If we can
continuously set $\alpha$ to zero, then the net number of chirals is
not affected.

So all in all, in the abelian case we have a fair amount of control.
The Betti numbers are computable when $\gamma = 0$ and should only
decrease when we turn on Higgs field monodromies starting from a
configuration with $\gamma = 0$. Further under such continuous
deformations, the net number of chirals is unchanged. In all cases,
thinking in the spectral cover picture, we can `close the loop' of
the boundary of a disk instanton by adding a segment along the zero
section, recovering the definition of Floer cohomology on the
cotangent bundle.

\newsubsection{Knot invariants}

Given a knot, link or graph $\Delta$ in $S^3$ with specified
$Gl(n,{\bf C})$ monodromies around it, there should be Lagrangian
three-cycle in $T^*S^3$ with boundary specified by $\Delta$. As
discussed in this paper, such a configuration defines an effective
$4d$ gauge theory, in fact it defines a whole ensemble of gauge
theories since we can embed the holonomy in different gauge groups.
Similar configurations have been considered before but focus on the
Wilson line correlation functions of the CS theory as knot
invariants. Here we see that there is another natural set of
observables which can be used as invariants, namely the $4d$
massless spectrum of the compactified gauge theory and its
superpotential (as a function of the massless chiral fields).
Presumably this encodes equivalent data, but it may provide a useful
reformulation.

\bigskip

\noindent%
{\it Acknowledgements}: We are indebted to B. Acharya, C. Beasley,
R. Donagi and N. Hitchin for discussion, and to Chris Beasley for
comments on the manuscript. The research of TP is partially
supported by NSF grant DMS-0700446 and NSF Research Training Group
Grant DMS-0636606.

\newpage

\newpage

\appendix

\renewcommand{\newsection}[1]{
\addtocounter{section}{1} \setcounter{equation}{0}
\setcounter{subsection}{0} \addcontentsline{toc}{section}{\protect
\numberline{\Alph{section}}{{\rm #1}}} \vglue .6cm \pagebreak[3]
\noindent{\bf Appendix {\Alph{section}}:
#1}\nopagebreak[4]\par\vskip .3cm}

\newsection{Some elements of Morse theory}
\seclabel{MorseReview}

We only summarize some of the basics here. See \cite{MilnorMorse}
for a classic account and \cite{Schwarz,Witten:1982im} for a more
modern perspective.

Consider a compact manifold $M$ and a function $f: M \to {\bf R}$.
The function $f$ is said to be Morse if all its critical points are
non-degenerate, i.e. the Hessian at the critical point has no zero
eigenvalues. We pick an auxiliary metric $g$ and define the gradient
flow
\be {\del \vec{x}\over \del \lambda} = -\nabla f \ee
We denote by $\phi_\lambda(x)$ the solution of this equation (the
gradient flow trajectory) which starts at $x$ at time $\lambda=0$.
Since any critical point is non-degenerate, up to coordinate
transformations the only invariant data of the Hessian is the number
of positive and the number of negative eigenvalues of the Hessian.
The number of negative eigen-values is called the Morse index of the
critical point.

Morse functions can be used to deduce the homotopy type of the
manifold $M$. Given a critical point $p$, we define the unstable
manifold of $p$ to be the set of points on $M$ that lie on a
gradient flow trajectory starting at $p$:
\be W_u(p) = \{ x \in M| \phi(x,-\infty) = p \} \ee
Similarly one may define the stable manifold $W_s(p)$, which
consists for the gradient flow trajectories ending at $p$. Then
$W_u(p)$ has the topology of a cell of dimension $i$, where $i$ is
the Morse index of $p$, and $M$ is the union of these cells
\be M = \cup_p W_u(p) \ee
To learn about the homotopy type, we have to understand how the
cells are glued together. We define the manifolds-with-boundary
\be M_a = \{ p\in M|f(p) \leq a\} \ee
The manifolds $M_a$ and $M_b$ are diffeomorphic if there are no
critical points in $f^{-1}[a,b]$. However if there is a critical
point, then the topology changes. For a critical point $c$ of Morse
index $i$, the manifold $M_{c+\epsilon}$ is homotopy equivalent to
$M_{c-\epsilon} \cup W_s(c)$. The proof is a local argument near
each critical point, see \cite{MilnorMorse} for details.

Unfortunately the above decomposition of $M$ does not necessarily
define a CW complex, and so the relation with the homology of the
manifold remains unclear. The missing condition was introduced by
Smale. The pair $(f,g)$ is said to be Morse-Smale if the stable and
unstable manifolds intersect each other transversally. This can be
shown to be satisfied for generic $(f,g)$.

Using such Morse-Smale pairs $(f,g)$, one may construct a homology
called the Morse homology. We define
\be C_i = \sum_{{\rm Crit}_i(f)} {\bf Z} p \ee
to be the free abelian group generated by the critical points of
Morse index $i$. The boundary operator is defined as follows. Given
two critical points $p,q$, we denote the moduli space of gradient
flow trajectores connecting $p$ and $q$ by $M(p,q)$:
\be M(p,q) = W_u(p) \cap W_s(q) \ee
When non-empty, the dimension of $M(p,q)$ is $m(p)-m(q)$, the
difference between the Morse indices. There is a natural action of
the real line ${\bf R}$ on this moduli space, given by rescaling the
parameter $\lambda$ that parametrizes the gradient flow trajectory.
Modding out by this rescaling, we define
\be n(p,q) = M(p,q)/{\bf R} \ee
whose dimension is $m(p)-m(q)-1$. If the Morse indices differ by
one, then $n(p,q)$ is zero dimensional and counts the number of
trajectories connecting $p$ and $q$. The moduli spaces $M(p,q)$ and
$n(p,q)$ come with natural orientations, and thus $n(p,q)$ counts
trajectories with a sign. We can define a boundary operator
\be \del: C_i \to C_{i-1}, \qquad \del p = \sum_{q \in {\rm
Crit}_{i-1}(f)} n(p,q) q \ee
The fact that $\del^2 = 0$ relies on an analysis of broken flow
lines, which lie on the boundary of the moduli space $M(p,r)$ with
$m(p)-m(r)=2$ and inherit their orientation from $M(p,r)$.

 \begin{figure}[t]
\begin{center}
            \scalebox{.3}{
               \includegraphics[width=\textwidth]{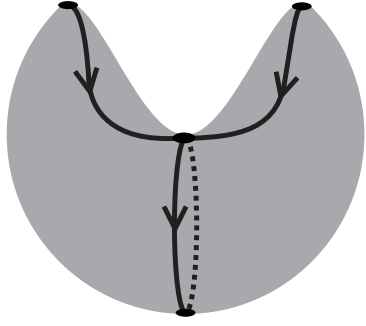}
               }
\end{center}
\vspace{-.5cm}
\begin{center}
\parbox{14cm}{\caption{ \it Gradient flow trajectories
connecting critical points of the height function on a deformed
two-sphere.}\label{MorseSphere}}
\end{center}
 \end{figure}

The correspondence with the ordinary homology of the underlying
manifold can be understood by establishing an isomorphism with
cellular homology, which is built from the free abelian groups
generated by the cells of a cell decomposition. We have already seen
how a Morse function gives rise to a cellular decomposition. To each
critical point of index $i$ we can associate an $i$-cell, namely the
associated unstable manifold, and the original manifold is precisely
the union of these cells. For Morse-Smale pairs, the boundary of
such an $i$-cell is contained in the skeleton built from the
$k$-cells with $k < i$. The unstable manifold $W_u(p)$ for $p \in
C_i$ gets attached to the unstable manifolds $\cup_q W_u(q)$ where
$q\in C_{i-1}$ runs over the critical points connected to $p$ by
gradient flow. The boundary operator of the cellular complex maps
each cell to its boundary, with an induced orientation. Similarly
the boundary operator of the Morse complex induces a boundary map on
the cell complex, since there is a one-one correspondence between
critical points and generators of the cell complex. It will probably
not come as a surprise that these two boundary operators are
equivalent.

Dually we can also define the Morse cohomology. This is more natural
than homology for physicists, because it describes properties of
functions on the underlying manifold, rather than the underlying
manifold itself (although these two points of view are of course
related by Poincar\'e duality). We write generators of $C_i$ as
bra-vectors, and define the dual as
\be C^i = \Hom(C_i,{\bf R}),\qquad  \delta:C^i \to C^{i+1}, \qquad
\vev{r|\delta p} = \vev{\del r|p}\ \  \forall\ \bra{r}\in C_{i+1}\ee
or more explicitly, up to a similarity transformation,
\be \delta \ket{p} = \sum_{q \in {\rm Crit}_{i+1}(f)} n(q,p) \ket{q}
\ee
Intuitively one may think of a generator $\ket{p}$ of $C^i$ as a
delta-function differential form localized at the critical point
$p$, with $i$ indices all of which lie along the unstable directions
at $p$. We are using the same numbers $n(p,q)$ as before, but the
Morse index is increasing instead of decreasing. Thus this is
isomorphic to replacing negative gradient flow with positive
gradient flow, or replacing $f\to -f$, which is the incarnation of
Poincar\'e duality in this context.

Another way to think about Morse cohomology was introduced by
Witten, and is closer to the M-theory/Yang-Mills picture. The
conjugation
\be d \to e^{tf} d e^{-tf} \ee
is a similarity transformation and induces an isomorphism between
the ordinary De Rham cohomology at $t=0$ and a deformed cohomology
as $t \to \infty$. By the arguments reviewed in section
\ref{MorseCohomology}, the deformed cohomology is identified with
the Morse cohomology. However since the two are related by a
similarity transformation, the betti numbers don't depend on $t$.

A well-known example is shown in figure \ref{MorseSphere}. Let us
apply Morse theory to the height function of the deformed two-sphere
shown in this figure. There are four critical points, two of them
with Morse index two, one with index one and one more with index
zero. However the boundary operator maps the saddle point to the
difference between the two maxima. Therefore the Morse cohomology is
given by
\be H^2(S^2,f) = {\bf R}, \qquad H^1(S^2,f)=0, \qquad H^0(S^2,f) =
{\bf R} \ee
which is of course the same as the ordinary cohomology of $S^2$.

\end{document}